\documentclass[reprint,aps,pra,superscriptaddress]{revtex4-1}
\usepackage{graphicx}
\usepackage{dcolumn}
\usepackage{bm}
\usepackage{float}
\usepackage{amsmath}
\usepackage{amssymb} 
\usepackage{xspace}
\usepackage{xcolor}
\usepackage{color,colortbl}
\usepackage{enumerate}
\usepackage[left=23mm,right=13mm,top=35mm,columnsep=15pt]{geometry} 
\usepackage[english]{babel}
\usepackage[utf8]{inputenc}
\usepackage{cancel}
\usepackage{mathtools}
\usepackage{adjustbox}
\usepackage[T1]{fontenc}
\usepackage{lipsum}
\usepackage{lineno}
\usepackage{cellspace}
\usepackage{multirow,bigstrut}
\usepackage[update,prepend]{epstopdf}
\usepackage[breaklinks=true,colorlinks=true,linkcolor=blue,urlcolor=blue,citecolor=blue]{hyperref} 

\usepackage{tikz,xcolor,hyperref}

\definecolor{lime}{HTML}{A6CE39}
\DeclareRobustCommand{\orcidicon}{%
	\begin{tikzpicture}
	\draw[lime, fill=lime] (0,0) 
	circle [radius=0.16] 
	node[white] {{\fontfamily{qag}\selectfont \tiny ID}};
	\draw[white, fill=white] (-0.0625,0.095) 
	circle [radius=0.007];
	\end{tikzpicture}
	\hspace{-2mm}
}

\foreach \x in {A, ..., Z}{%
	\expandafter\xdef\csname orcid\x\endcsname{\noexpand\href{https://orcid.org/\csname orcidauthor\x\endcsname}{\noexpand\orcidicon}}
}

\definecolor{DarkGray}{rgb}{.0,.0,.0}
\definecolor{BrightGray}{rgb}{.85,.85,.85}
\definecolor{LightGray}{rgb}{.96,.96,.96}
\definecolor{LightCyan}{rgb}{0.88,1,1}

\newcommand{\be}{\begin {equation}}
\newcommand{\ee}{\end {equation}}

\begin{document}

\title[Controlled transition to different proton acceleration regimes]{Controlled transition to different proton acceleration regimes: near-critical density plasmas driven by circularly polarized few cycle pulse}

\author{Shivani Choudhary\orcidC{}}
 \affiliation{ELI-ALPS, ELI-HU Non-Profit Ltd., Wolfgang Sandner utca 3., Szeged 6728, Hungary}

\author{Sudipta Mondal\orcidB{}}
\affiliation{ELI-ALPS, ELI-HU Non-Profit Ltd., Wolfgang Sandner utca 3., Szeged 6728, Hungary}%
\author{Daniele Margarone}
\affiliation{ELI Beamlines Center, Institute of Physics, Czech Academy of Sciences, Za Radnic\'{i} 835, 252-41 Doln\'{i} B\v{r}e\v{z}any, Czech Republic}

\author{Subhendu Kahaly\orcidA{}}%
 \email{subhendu.kahaly@eli-alps.hu}
 \affiliation{ELI-ALPS, ELI-HU Non-Profit Ltd., Wolfgang Sandner utca 3., Szeged 6728, Hungary}
  \affiliation{Institute of Physics, University of Szeged, D\'om t\'er 9, H-6720 Szeged, Hungary}
%

\date{\today}

\begin{abstract}
We investigate the different facets of ion acceleration by a relativistically intense circularly polarized laser pulse interacting with thin near-critical density plasma targets. Our simulations establish that plasma density gradient and laser frequency chirp can be controlled to switch the interaction from the transparent to the opaque regimes of operation. This enables one to choose between a Maxwellian like ion energy distribution with a cut-off energy, in the relativistically transparent regime, or a quasi-monoenergetic spectrum, in the opaque regime. We subsequently demonstrate that a double-layer multi-species target configuration, can be effectively utilized for optimal generation of quasi mono-energetic ion bunches of a desired species. We finally demonstrate, the feasibility of generating mono-energetic proton beams with energy peak at $\mathcal{E}\approx20\sim40$ MeV with a narrow energy spread of $\Delta \mathcal{E}/\mathcal{E}\approx18-28.6\%$ confined within a divergence angle of $\sim 175$ millirad at a reasonable laser peak intensity of $I_{0}\simeq 5.4\times 10^{20}\,  \mathrm{W/cm^2}$.
\end{abstract}

\maketitle

\section{Introduction}
The technological advancement in targetry technology and a growing interest in near-critical density plasma \cite{Fedeli2018b,Ji2018}, both in utilizing gas-based \cite{Sylla2012,Lifschitz2014,Kahaly2016,Henares2019} as well as solid-based \cite{Yogo2008,Pazzaglia2020,Martinez2020,Wang2021} interaction, paved a new regime of particle acceleration in relativistic intensity laser-plasma interaction. The advent of powerful ultrashort multi-cycle \cite{Danson2019,Radier2022} and few-cycle laser facilities worldwide \cite{Kuhn2017,Charalambidis2017} along with sophisticated beamlines \cite{Mondal2018,DanieleM2018} makes such experiments feasible in the near future. Over several decades, the quest to achieve monoenergetic proton beams of MeV energy has influenced the growth in the field of laser-based particle acceleration research \cite{Macchi2013_RMP, Schreiber2016}. 
The interaction of ultra-intense, ultra-short pulses with the plasma target leading to the generation of relativistic energy particles with enormous potential for applications in materials \cite{Passoni_2019} and medical sciences \cite{Bulanov2002_PPR,Bulanov2014_PU,Karsch2017_TandF}, fusion schemes \cite{Weng2018_SD, Roth2001}, industrial applications \cite{Barberio2017_nature} and as a neutron source\cite{Roth2013_PRL}.

A diversity of ion acceleration processes have already been identified and demonstrated. For example, Target Normal Sheath Acceleration (TNSA) \cite{Snavely2000_PRL, Wilks_2001, Mora2003_PRL} has been the most widely studied and investigated ion acceleration mechanism. TNSA predominantly occurs when a high intensity laser pulse interacts with an opaque solid density thick foil target producing a proton beam with maximum proton energy in the range of several tens of MeV \cite{Hegelich2006_NAT,Ogura2012OL}. In this process, the hot electrons generated during the laser interaction transport through the target and exit from the rear end, creating a sheath electric field that accelerate the protons and other ions present at the back end layer of the target along the direction of target normal \cite{Passoni_2010}. The constraint with TNSA is mainly the broad energy spectrum that makes the use of such beams for societal application very challenging if beam conditioning (energy selection) is not implemented. Energy selection, on the other hand, reduces the ion beam flux on a sample (i.e. a biological one) as in for cancer therapy \cite{Kroll2022}.

An intense ultrashort laser pulse also imparts radiation pressure during the interaction, which is a function of the reflectivity of the relativistically exited foil target. The complete transmission of ultra-intense laser light from the plasma surface results in zero radiation pressure, while it maximises in the case of complete reflection. In a typical laser-plasma interaction at relativistic intensity, the radiation pressure can reach nearly giga-bar ($10^9$ bar) level \cite{Chou2022, Natsumi2018_Nat} and under optimal conditions this can favourably accelerate ions. Radiation Pressure Acceleration (RPA) has been demonstrated to operate in two different acceleration regimes, Hole-Boring (HB) RPA \cite{Robinson2006_PRL, Robinson2008_NJP, Natsumi2018_Nat} ($\mu$m thick targets), and Light-Sail (LS) RPA \cite{macchi_ls,macchi_njp} (for nm thin targets), with distinct features in the resulting ion spectra.  RPA has its own limitations: such as LS-RPA works well only at ultrahigh intensities ($\gg 1\times10^{21} W/cm^2$) and HB-RPA requires long pulse ($\sim$ps), high energy lasers since a relatively long time and large energy in the pulse is required to effectively drill a hole into the target \cite{Natsumi2018_Nat} and to accelerate ions at the target front side through RPA. The maximum proton energy from target in RPA regime observed is about $\mathrm{48\,MeV}$ \cite{Ma2019_PRL}.

Light reflection plays a crucial role in the relativistic interaction. In non-relativistic plasma, an incident laser pulse with frequency $\omega$ lower than the plasma frequency, $\omega_p = \sqrt{n_e e^2/\epsilon_0 m_e}$ (where $n_e$ is the electron density of the target plasma, $\epsilon_0$ is vacuum permittivity, $e$ and $m_e$ are the electron charge and mass) is reflected. The critical plasma density, $n_c = m_e \omega^2 \epsilon_0 / e^2$, at which the plasma frequency equals the wave frequency marks the transparency threshold and for $n_{e}>n_{c}$, the plasma is described as overdense. For relativistically intense lasers, the plasma electrons are accelerated by the laser field lowering the effective critical density by a factor of $\langle\gamma\rangle$ \cite{Sprangle1990_PRL, Fedeli2017,Natsumi2018_Nat,Pazzaglia2020}, which is the average Lorentz factor if the electrons in the reflecting layer increasing the transparency threshold to, $n_{c}^{rel}\approx \langle\gamma\rangle n_{c}$ \cite{Fedeli2017}. Thus, the plasma becomes relativistically underdense for, $n_{e}<n_{c}^{rel}$. This is the relativistically induced transparency (RIT) regime that optically switches opaque plasma to transparent, enabling light propagation \cite{Cattani2000_PRE, Goloviznin2000_POP, Siminos2012_PRE}. In the recent years, several articles have reported efficient proton acceleration from the ultra-thin nm-scale targets in this regime \cite{Juan2017_POP,Sahai2013_PRE, Poole2018_iop, Palaniyappan2012_NAT, Higginson2018_nature,Singh2022}. \newline

In this article, via a series of fully relativistic 2D plane wave (one dimension in coordinates and three dimensions in velocities) particle in cell (PIC) simulations, over a wide range of laser and target parameters, we demonstrate that one can transit between two-different ion acceleration mechanisms, namely, Relativistic Induced Transparency (RIT) and Radiation Pressure Acceleration (RPA). These regimes of ion acceleration proposed in this study do not necessarily require PW-class lasers but potentially can be achieved using sub-PW peak power and moderate ($\sim 10^{20} W/cm^2$) laser intensities. These parametric variations provide a route to effectively tune the spectral characteristics of the generated ions from the target. We investigate the influence of peak plasma density, target thickness, plasma density gradient, laser temporal chirp, and laser focal spot size effects on the ion acceleration process from the near-critical density plasma. We identify conditions that can effectively accelerate quasi mono-energetic ions with narrow energy spread. The observations may potentially be investigated experimentally at ELI-ALPS \cite{Kuhn2017,Charalambidis2017,Mondal2018} and ELI beamlines \cite{Schillaci2022} in the near future. Additional 2D-PIC simulations give a clear indication on how target configuration (Double layer target) can provide the mono-energetic ion bunches of the species of choice depending upon the ion acceleration mechanisms and sheds light on the mechanism of quasi mono-energetic ion acceleration.

The work done in this article is structured as follows. We discuss the ion acceleration from the foil near-critical density targets with step-like density profile and bring out the distinctive features of the processes in the transparent and the opaque regimes both in real space and the phase space, in section \ref{sec:foil}. In section \ref{sec:steptarget} we summarise our observations on target thickness and plasma density effects on the transition to different proton acceleration regimes, and investigate the robustness of the correlation between the optical transparency and the nature of proton energy spectral shape. We discuss the influence of an exponential density ramp in front of foil target, a feature that can be all optically controlled in experiments with a pre-pulse, in section \ref{sec:gradient}. In section \ref{sec:chirp_main}, we explore and summarise the effect of temporal chirp on the ion spectral characteristics for different target species and show that preferential acceleration of a chosen ion species can be done using a double layer target in section \ref{sec:tuning}. Finally, in section \ref{sec:2d}, we discuss the effect dimensionality and of tight focusing the ion energies from the double layer target configuration and probe the quasi mono-energetic ion acceleration deeper.
\section{Proton acceleration in step like sub-wavelength scale near critical density foil}
\label{sec:foil}

 \begin{figure}[t!]
    \vskip -0.35cm 
    \begin{center}
    \includegraphics[scale=0.85]{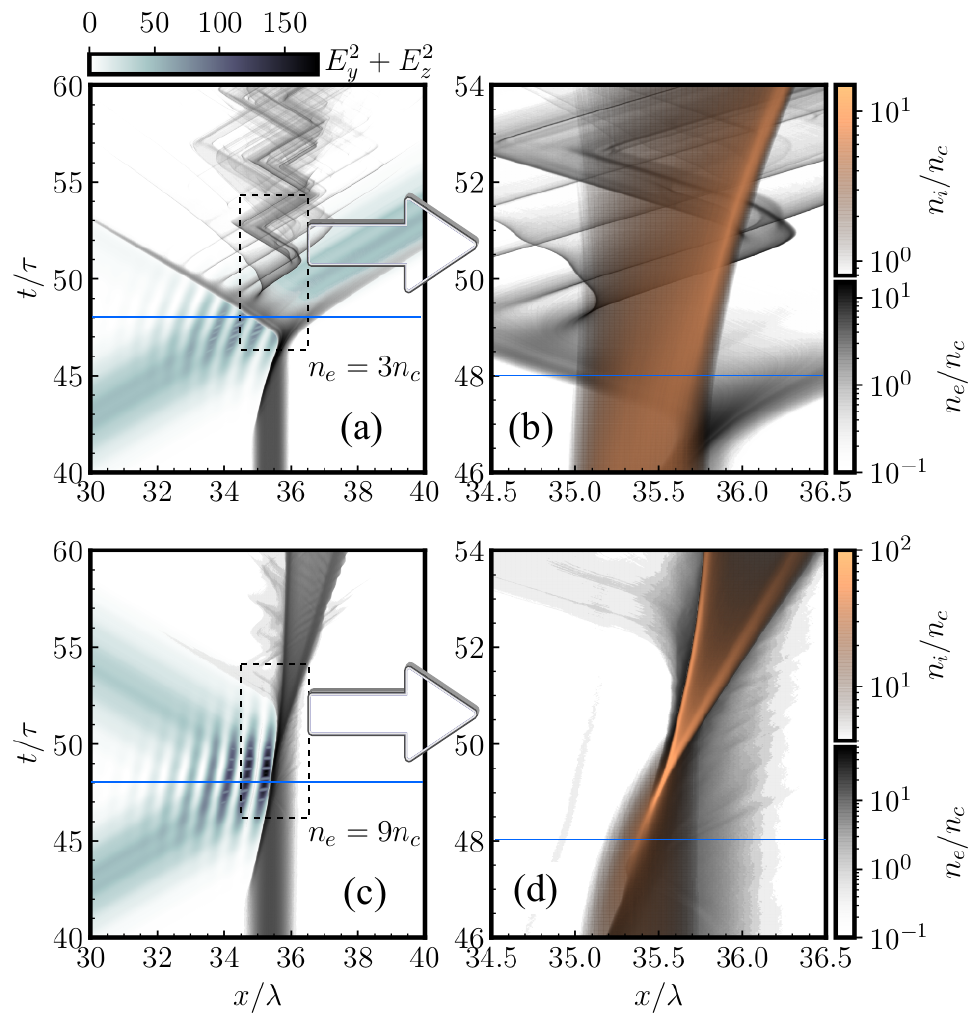}
    \vskip -0.45cm 
    \caption{Spatio-temporal dynamics of an intense few-cycle circularly polarized electric field driven thin slab target with thickness in the sub-$\lambda$ regime (in this case foil of thickness $d= 0.75\lambda$). The spatio-temporal evolution of the laser intensity (${E_{y}^2+E_{z}^2}$ presenting the incident, reflected and transmitted light), electron density ($n_{e}$) and ion density ($n_{i}$), is depicted for two different initial plasma densities $n_{0}$ representative of the two distinct scenarios: (a)-(b) $n_{0}$=$n_{e}$=3$n_{c}$ representing light transmission and (c)-(d) $n_{0}$=$n_{e}$=9$n_{c}$ showing complete reflection of light. The solid horizontal lines mark the instant $t=48\tau$ when the peak of the pulse envelope interacts with the target surface. (b) and (d) represents a zoomed view of the dotted boxes demarcated in (a) and (c) respectively, capturing the distinct signatures of the ion and electron dynamics in the two different regimes during the relativistic interaction. Both the targets are initially overdense and reflect the laser (since $n_{0}>n_{c}$), which interferes with the incident pulse forming the standing wave pattern in front of the target (see the intensity fringes on the front side of the target). (a) Near the peak of the pulse, the target becomes optically thin (dynamically underdense), pushing out electrons from the surface, initiating complex electron dynamics in the ion background and expanding the ion density distribution shown in (b). (c) The target retains overdense character within the laser pulse duration leading to in sink motion of the electron and ion density peaks shown in (d). We have used a Gaussian laser pulse with normalized peak laser pulse amplitude $a_0$ = 20 (as in eq. \ref{gp}) for both cases.}\label{ey2ez2ne}
    \end{center}
     \vskip -.75cm
\end{figure}

The interaction of a relativistic intensity laser pulse with a semi-infinite plasma and subsequent plasma dynamics can be modeled accurately, to a large extent, using 1D cold fluid model \cite{kaw}. In the reflection region (vacuum side), this theory provides the stationary solution at the vacuum plasma boundary by balancing the ponderomotive and the electrostatic forces. In the plasma side, the laser ponderomotive force pushes electrons inside the plasma to a new position leaving the immobile ions behind, consequently, a strong charge separation field is created which peaks at the new electron position. However, at the interaction condition, when most of the electrons escape from the target, the laser pulse starts to propagate deeper inside, thus, the electron dynamics become more complex and the validity of cold-fluid approximation ceases to exist. Therefore, PIC simulations are required to understand the behavior of interaction at this regime.

The laser electric field $\mathbf{a}$ for a Gaussian laser pulse with elliptical polarization can be defined as follows,
\be \mathbf{a}(\eta) = \frac{a_0 f(\eta)}{\sqrt{1+\epsilon^{2}}}\bigg [\mathrm{cos} [\phi(\eta)] \mathbf{e_y} + \epsilon~ \mathrm{sin}[\phi(\eta)]\mathbf{e_z}\bigg] \label{gp}\ee
where, the laser electric field  $\mathbf{a}(\eta)$ is in units of $m_{e}\omega c/e$, $f(\eta) = exp[-4 ln (2) \frac{\eta^2}{\tau_{FWHM}^2}]$ is the normalised pulse envelope function,  $\phi(\eta) = 2\pi \eta$ is the phase function and $\eta = t-x$ is the propagation coordinate (where the space and time coordinates are normalised with respect to the time period $\tau$ corresponding to the central carrier frequency and the laser central wavelength $\lambda$ respectively). $\tau_{FWHM}$ is Full-Width-at-Half-Maxima (FWHM) of the time dependent intensity envelop and $a_{0}$ is normalized laser pulse amplitude. Here, $m_e$ and $e$ are the mass and charge of the electron. $\epsilon$ is the ellipticity parameter which can vary from -1 (left-handed circular polarization) to 0 (linear polarization) to +1 (right-handed circular polarization), where the sign determines the helicity and the magnitude defines the ellipticity in general. We kept $\epsilon = -1$ for all the cases studied under the present investigation.

Furthermore, using 1D model equations, one can obtain the threshold limit for the self-induced transparency and to differentiate between the regions of transparency and opacity for semi-infinite overdense plasma target in relativistic framework \cite{Cattani2000_PRE, Goloviznin2000_POP}. In a more rigorous manner, using fully relativistic PIC simulations one can obtain the percentage of laser pulse energy reflection and transmission from the target surface. In our study we follow the second procedure, while conceptually benefiting from comparing the results with the expectations from the model. A constant parameter $\xi_0$ is taken into account which defines the normalized surface density of the target \cite{Vshivkov1998_POP}, and is proportional to the product of target density $(n_e)$ and thickness $(d)$ of the thin plasma slab and is given by, $\xi_0 = \pi \bigg(\frac{n_e }{n_c}\bigg)_{\omega_{0}}\frac{d}{\lambda}$. In case of moderate intensities ($a_0 < 1$), $\xi_0 < 1$ corresponds to transparency regime. Whereas, for ultra-intense laser pulses ($a_0 >1$), in the non-linear regime, the condition of transparency is achieved when $\xi_0 > 1$ and $a_0 \geq \xi_0$ (or $a_{0}/\xi_{0}\geq 1$) \cite{Vshivkov1998_POP}. 

In order to have a clear understanding about the region of transparency and opacity, we have used 2D plane wave simulations with fully relativistic PIC simulation code LPIC++ \cite{lpic} with modifications including in both target conditions such as an introduction an of additional layer, exponential ramp, as well as implementing Gaussian profile and chirp function in the laser. In order to investigate the role of peak plasma density $n_{e0}$ for a given laser central frequency and foil thickness, we consider a quasi-neutral plasma foil, $n_{e}(x)= n_{e0}[H(x-x_{1})-H(x-x_{2})]$, where $H$ is the Heaviside step function and $d=x_{2}-x_{1}$ is the target thickness. 

A circularly polarized (CP) laser pulse with normalized amplitude $a_0$ = 20 ($I_0\simeq 5.4\times 10^{20}\,  \mathrm{W/cm^2}$), field envelope FWHM = 5 cycles is normally incident on the sharp gradient plasma slab of thickness 0.75$\lambda$ from the left side of the simulation domain. The simulation domain is of 100$\lambda$ with 140 cells per wavelength and 1500 particles per cell. The laser and target parameters are taken in dimensionless units. The space and time are normalized as  x/$\lambda$ and t/$\tau$ respectively, where  $\lambda = 1\,\mathrm{\mu m}$ is laser wavelength and $\tau = \lambda$/c. The electron density is normalized with respect to critical density $n_{c}$ and fields are normalized as $e E/m_e \omega c \rightarrow E$. The plasma slab is located from 35$\lambda$ $< x <$ 35.75$\lambda$ having density of 3$n_c$ which is considered as nominally overdense, for the intensity of $\sim \mathrm{5.4\times 10^{20}\, W/cm^2}$ and hence calculates $\xi \sim$ 7 and therefore $\frac{a_0}{\xi} > 1$ satisfying the transparency condition. On the other hand, $n_e$ = 9$n_c$, is a fully overdense region for the given intensity and results in $\xi \sim$ 21.195, where the ratio $\frac{a_0}{\xi} \lesssim 1$, therefore tapping into the total reflection region \cite{macchi_ls, Choudhary2016_EPJD}. Such kind of near-critical densities are used in experiments from, foam targets \cite{Willingale2011_POP}, cryogenic hydrogen jet target \cite{obst2017_sr} and cryogenic solid hydrogen target \cite{Polz2019_sr}

In Figure \ref{ey2ez2ne}, we present the spatio-temporal profile of laser intensity, electron, and ion density for the given laser and target parameters. Figure \ref{ey2ez2ne}(a) and (c) represent the two dynamically opposite cases with nearly (a) $\sim 90 \% $ (c) $\sim 0 \% $ transmission of laser pulse through the target. Therefore, we refer to these two regions of ion acceleration, with the nominally underdense region as in \emph{relativistic transparency} (a, b) and overdense region as in \emph{opacity} (c, d). 

When a circular polarized (CP) laser pulse is incident on the target, due to the absence of $J\times B$ heating phenomenon within the interaction process, in relativistically transparency region \ref{ey2ez2ne}(a), sufficient number of electrons escapes the target, thus leaving the target to be underdense. As the laser pulse propagate through the target, the electrons move in the vicinity of the laser field in both forward (target rear) and backward (target front) direction. Whereas, ions being heavier in mass, undergoes a marginal expansion around the target surface (from 35$\lambda$ to 35.75$\lambda$), mostly in the forward direction as shown in Figure \ref{ey2ez2ne}(b). The electrons which are removed from the target ion background create a charge separation at the interface generating a strong electrostatic field. As a result, the electrons are pulled back to the target surface towards the ions and oscillate around the interface of the target. A magnified illustration of this process is shown in Figure \ref{ey2ez2ne}(b) with ion and electron density profiles. The horizontal cyan lines in Figure \ref{ey2ez2ne} demarcates the time at which the peak of the laser field interacts with the target. As seen in \ref{ey2ez2ne}(a), as the rising part of the laser field interacts with the target, the ponderomotive force imparted by the laser field compresses the electron into the ion background, eventually pushing out part of the electrons near the peak of the laser field leading to light transmission.  Afterwards rest of the electrons execute back and forth oscillations in the ion background leading to expansion of the ions as seen in \ref{ey2ez2ne}(b). 

The interaction in the opaque regime ($n_e = 9n_c$) is presented in Figure \ref{ey2ez2ne}(c,d). In Figure \ref{ey2ez2ne}(c), we observe a completely different electron dynamics.
Due to the interaction of CP laser pulse with overdense target, most of the laser pulse is reflected. Therefore, all the electrons are compressed inside the target and are piled up at the rear side of the target leading to the formation of two electron bunches at a later time. In order to have a clear view, we present a zoom into the spatio-temporal profile of electron and ion density during the interaction in Figure \ref{ey2ez2ne}(d). We observe that, the electrons and ions position closely overlaps with each other and co-propagate together, as shown in ref. \cite{Ji_2018NJP}. At this point we would like to note that the partial transmission and reflection of the light field during the interaction is well captured in the colormap presenting the intensity envelop in Figure \ref{ey2ez2ne}(a) and (b). A standing wave pattern is clearly visible in the target front side, representing the interference of the incident and reflected light fields and hence persisting only as long as the light reflection persists during the interaction. As seen in Figure \ref{ey2ez2ne}(a), near the cyan line, the target becomes partially transparent, reducing thereafter the contrast in the standing wave pattern.
 \begin{figure*}[t!]
    \vskip -0.35cm 
    \begin{center}
    \includegraphics[scale=0.45]{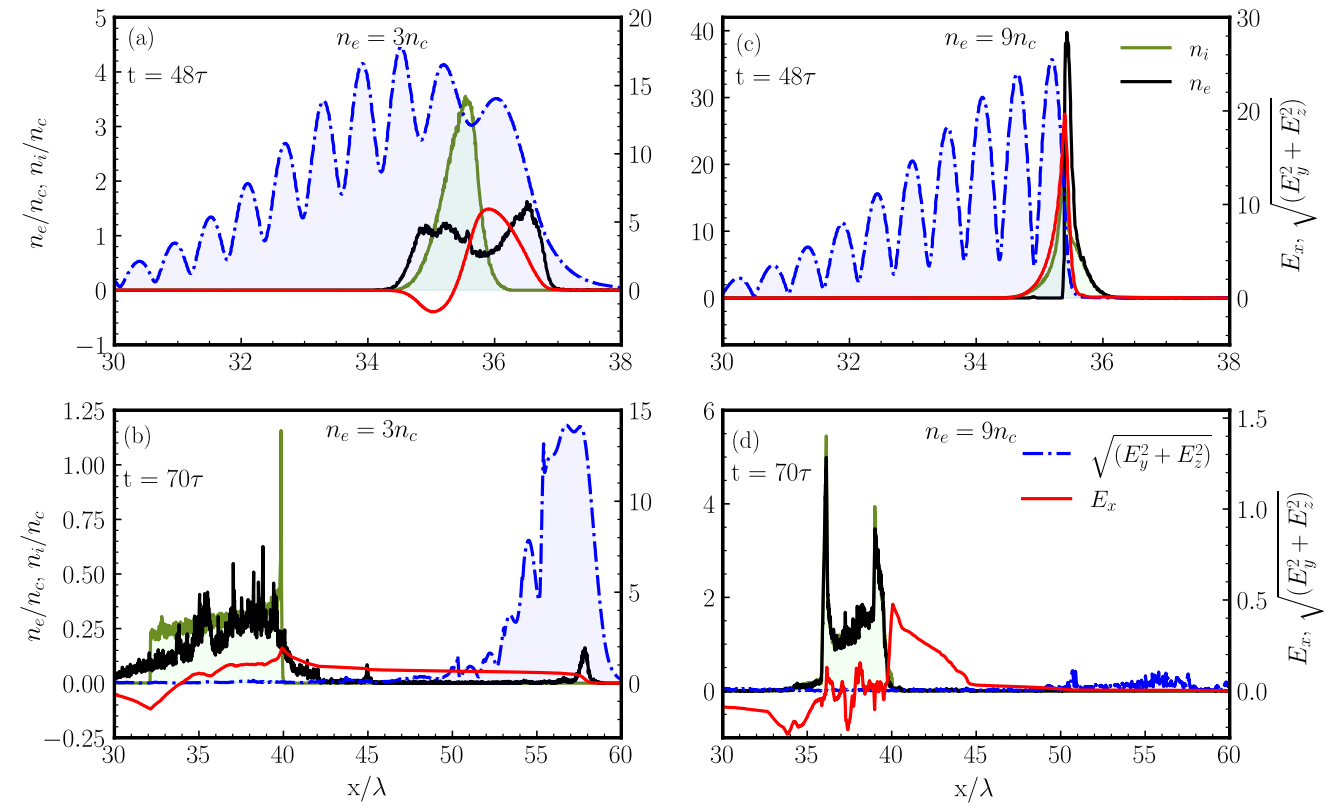}
    \vskip -0.45cm 
    \caption{Temporal snapshot depicting the profiles of the driver laser field magnitude (dashed blue line representing the incident, reflected and transmitted fields),  longitudinal electric field (red solid line representing the accelerating field), ion density (green solid line) and electron density (black solid line) for the two scenarios: (a)-(b) at $n_e$ = 3$n_c$ (relativistically transparent regime) and (c)-(d) at $n_e$ = 9$n_c$ (overdense regime). For each case the fields and densities are plotted at two different time instants: (a)-(c) at a time $\mathrm{t=48\, \tau}$, when the peak of the laser pulse is interacting with the target and (b)-(d) at latter time $\mathrm{t=70\,\tau}$, long after the driving field has ceased to interact with the target. Near the peak of the interaction, the lower initial density target already starts to transmit the incident laser pulse which can be seen in the blue shaded part on the right side of the target in (a) and is clearly captured after the interaction in the laser pulse co-propagating to the right along with the electron bunch as seen in (b). No such transmission can be witnessed in (c) and (d) (pay attention to the significantly small scale used for plotting the magnitude on the right axis for the radial electric field presented in (d)).\label{fieldens}}
    \end{center}
\end{figure*}
To discuss the underlying physics behind the ion acceleration in these two specific regions of interest, we present the transverse and longitudinal electric fields along with the ion and electron density profiles in Figure \ref{fieldens}, for the above mentioned laser intensity and target density cases. Here, we have considered the two specific time instants, one at t=48$\tau$, when the peak of the driver pulse interacts with the target and, other at t=70$\tau$ which we say as post-interaction. During these two time instances, we analyze the behavior of all the field and density components.

In the relativistic transparent region, at t=48$\tau$ in Figure \ref{fieldens}(a), the laser pulse energy starts to propagate through the target with limited laser energy reflected. This eventually leads to the formation of a standing wave pattern of low contrast at the front side of the target as mentioned before.
As discussed in the context of Fig \ref{ey2ez2ne}(a,b) due to the formation of the electron bunches on both sides of the target, a bipolar electrostatic field (5 arb.u.) is generated (red solid line) at t=48$\tau$ which is the signature of target expansion on both the side.
At a similar moment, a completely different behaviour for the opaque regime (Figure \ref{fieldens}(c)) is observed, which exhibits a unipolar longitudinal field. Since, in this case, a substantial amount of laser energy is reflected from the target front surface, and this leads to the formation of standing waves with high contrast. Therefore, the electrons receive a laser push in the forward direction in the form of radiation pressure and move inside the target, followed by the slowly moving ions. Thus, creating a charge separation only at the rear side of the target \cite{Guerin_1996POP} and hence, resulting in a unipolar electrostatic field (20 arb.u.), as seen in the red curve in Figure \ref{fieldens}(c). 

At a post interaction time (t=70$\tau$), we observe a strong persistent longitudinal electrostatic field in the relativistic transparent region as shown in Figure \ref{fieldens}(b). In this case, the laser pulse is already propagated through the target and a fraction of the electron population is expelled from the target to make it positively charged. This results in the ions to expand under its coulomb repulsion. A small fraction of electrons  with very high energy nearly co-propagate with the transmitted laser pulse (can be seen at 57$\lambda$) leading to the formation of a strong electrostatic field (2 a.u.). Since, the electrons escape from the target in both directions, therefore it results in the generation of a negative electrostatic field, at the target front.

In the case of overdense target, as seen in Figure \ref{fieldens}(d), at the later time instant of $70 \tau$, when complete laser pulse has interacted with the target, the ions and electrons move nearly together forming an overlapping double peak structure in space. This results in generation of a weak electrostatic field  $\sim$0.5 a.u. These bunches slowly expand in the target forward direction (rear side) under radiation pressure acceleration. This shows that the laser pulse transfers its momentum to the electron and ions and lets them evolve under this momentum transfer \cite{Robinson2009_PPCF,Robinson2012_PPCF}.
Similar to the cold fluid model,  the position of the maximum electrostatic field is at the minimum of the electron density, as shown in Figure \ref{fieldens}(c) at the peak interaction of the pulse. At the later instants ($70\tau$), the position of the peak electrostatic field cannot be correctly interpreted from the cold fluid model. Since ions being mobile and electron bunches leaving and re-entering the target gives rise to this difficulty, as mentioned in ref. \cite{Cattani2000_PRE}.

 \begin{figure*}[t!]
    \vskip -0.35cm 
    \centering
        \includegraphics[scale=0.75]{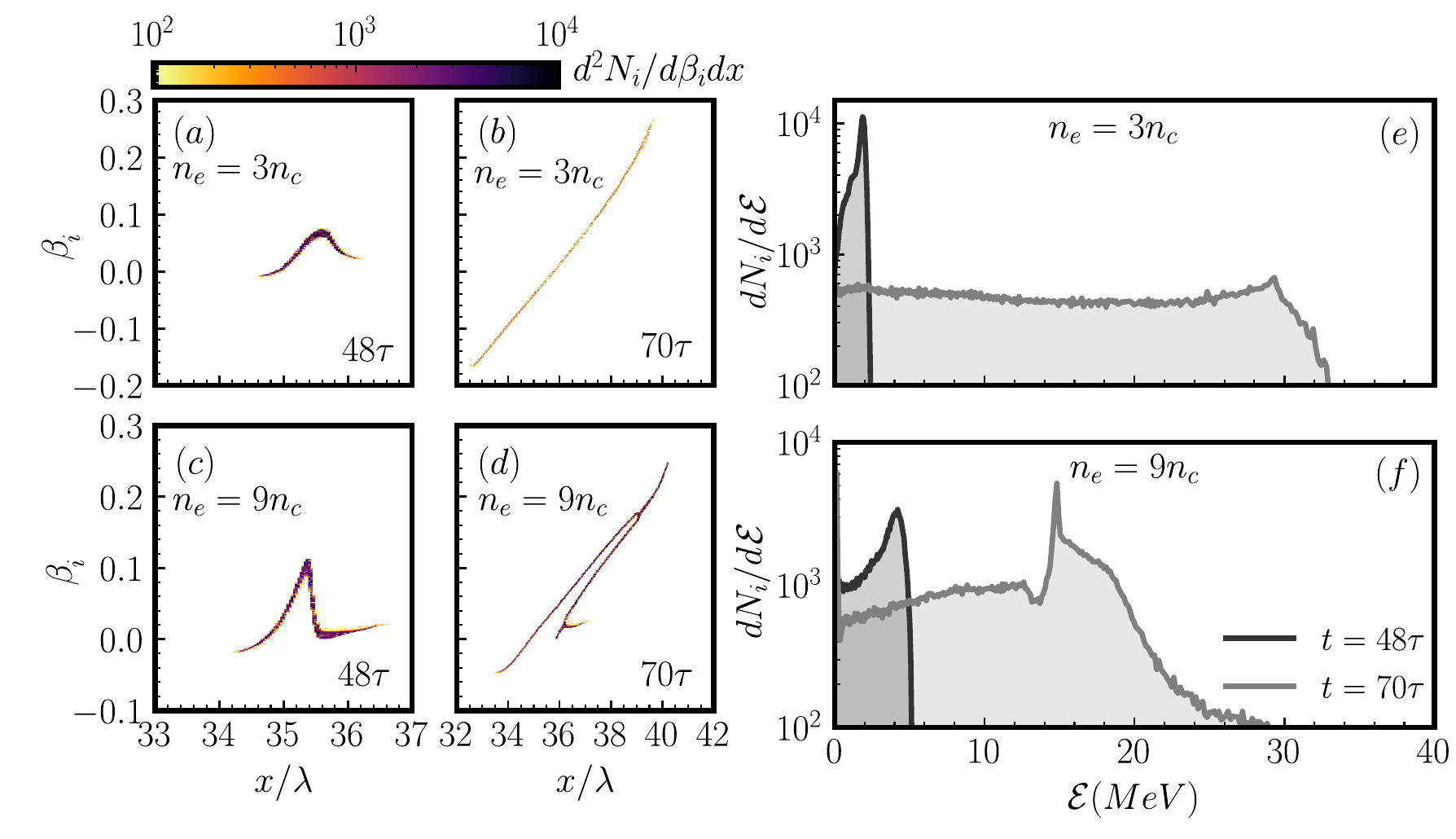}
    \vskip -0.45cm 
    \caption{Temporal snapshots of ion phasespace distribution and the corresponding energy spectra. (a)-(b) Ion phasespace distribution in the blow-out regime ($n_e = 3n_c$) of interaction at two different time instants.  (c)-(d) Ion phasespace distribution in the opaque regime ($n_e = 9n_c$) of interaction at the same time instants. At $\mathrm{t=48\, \tau}$ the laser pulse peak is interacting with the target. $\mathrm{t=70\, \tau}$ signifies an instant long after the driving field has ceased to interact with the target. The color bar in (a)-(d) represents the number of macroparticles $d^{2}N_{i}/d\beta_{i}dx$ accelerated in the laser propagation direction with the velocity ($\beta_{i}$), normalized with speed of light in vacuum ($c$), per unit bin in the phasespace. (e)-(f) The ion energy spectra $dN_{i}/d\mathcal{E}$ corresponding to the interactions represented in (a)-(d). Both the targets behave in a reflective manner until near the peak of the laser pulse envelop reached at $\mathrm{t=48\, \tau}$. At the peak of the driving field the foil with $n_e = 3n_c$ becomes transparent, whereas the one with $n_e = 9n_c$ remains reflective during the whole relativistic interaction. The two different behaviours, representative of the two different regimes of operation, are captured very well in this figure. During the reflective regime the phase space velocity distribution presented in (a),(c) and (d) demonstrate similar folded shape and peaky features which becomes significantly different in (b) once the interaction enters the transparent regime. Similar behaviour unravels their signatures within the ion energy spectra. For the reflective part of the interaction (dark-shaded curves in (e),(d)) the ion energy spectra show distinct quasi-monoenergetic behaviour which is retained even after the interaction is over as seen in the light-shaded curve (f), but erodes away once target enters transparent regime even partially as seen in the light shaded curve in (e).\label{phasespace}}
\end{figure*}
We have presented earlier, in Figure \ref{ey2ez2ne}(a) and elaborated in Figure \ref{fieldens}(a) for the case of the target with peak electron density $n_{e}=3n_{c}$, it becomes transparent near the peak of the laser field. On the other hand, in the case of the target with peak electron density $n_{e}=9n_{c}$ (as presented in Figure \ref{ey2ez2ne}(c) and Figure \ref{fieldens}(c)) the laser field fails to pass through the target even at the peak of the field envelop. In order to investigate the ion dynamics in more detail and to identify the differences between the regimes of transparency and opacity we now analyse the phase space data and look into the behaviour of the ions in above mentioned cases in Figure \ref{phasespace}. The ion velocity in the direction of laser incidence (along decreasing $x$), in units of $c$ is represented by $\beta_{i}$ in the colormap, where a positive $\beta_{i}$ in Figure \ref{phasespace} indicates ion velocity into the target (from left to right in Figure \ref{ey2ez2ne} and Figure \ref{fieldens}). Before the peak of the laser pulse interacts with the target, in both the cases, the target remains predominantly reflective, as is evident from the relatively high contrast of the interference fringes in the front side of the target (Figure \ref{fieldens}(a,c)). From the beginning of the interaction, upto this point in time, the electrons face the increasing Lorentz push from the laser field and gain in energy which is self consistently transferred to the ions via the plasma charge separation field. As is evident in Figure \ref{phasespace}(a), for the target which eventually becomes transparent at the peak-interaction (48$\tau$), the ions gain in forward momentum showing a $\beta_{i}$ peak located at $x=35.75\lambda$ (which is inside the initial target surface at $x=35\lambda$) in the phase space velocity distribution. We note here that the asymmetric bipolar charge separation field $E_{x}$ (the different values of the negative and positive peaks in the red curve in Figure \ref{fieldens}(a)) leads to the asymmetric ion velocity distribution in Figure \ref{phasespace}(a), being skewed towards the laser propagation direction. Long after the interaction (at $70\tau$) in the transparent target, as shown in the phase space distribution in Figure \ref{phasespace}(b), we observe that ions have expanded in both the direction (front and rear), with a slower ion expansion at the front side of the target (having a velocity cutoff near $\beta_i \sim 0.18$), than the rear side (with velocity cut off near $\beta_i \sim 0.27$). In this case, since the ions undergoes the Coulomb repulsion, it expands in both the direction (front and rear) as in Figure \ref{ey2ez2ne}(b).

Here we note that, if the features of the ion velocity distribution are closely linked with the regimes of transparency during the interaction, at this point, one may expect that until $t=48\tau$ both the targets should show qualitatively demonstrate similar ion phase space behaviour. This behaviour is fully corroborated by the similar features being observed in the opaque region, as shown in Figure \ref{phasespace}(c), near the peak of the pulse. However, there are few qualitative differences with respect to the scenario presented in Figure \ref{phasespace}(a). Firstly, in Figure \ref{phasespace}(c) we observe that at $t=48\tau$, the peak of the $\beta_{i}$ distribution is located at a higher value compared to the previous case. Secondly, in the case of $n_{e}=9n_{c}$, the asymmetric phase space ion velocity distribution is more skewed. Both these points can be well understood by looking at the $E_{x}$ field profile presented in Figure \ref{fieldens}(c), which shows a higher value of peak field and a sharper field profile inside the target than in the front side. Hence the ions show some expansion at the target front, but are much better piled up within target thickness, showcasing the build up of charge as shown in Figure \ref{phasespace}(c). At a longer time delay after the interaction ($t=70\tau$), for the opaque region (Figure \ref{phasespace}(d)), we observe two distinct ion velocity distributions, where one ion bunch accelerates from the front surface of the target, another ion bunch accelerates from the rear surface of the target. Eventually, both ion bunches merge to form a single ion distribution at $x$ = 39$\lambda$.
 \begin{figure*}[t!]
    \vskip -0.25cm 
    \centering
        \includegraphics[scale=0.90]{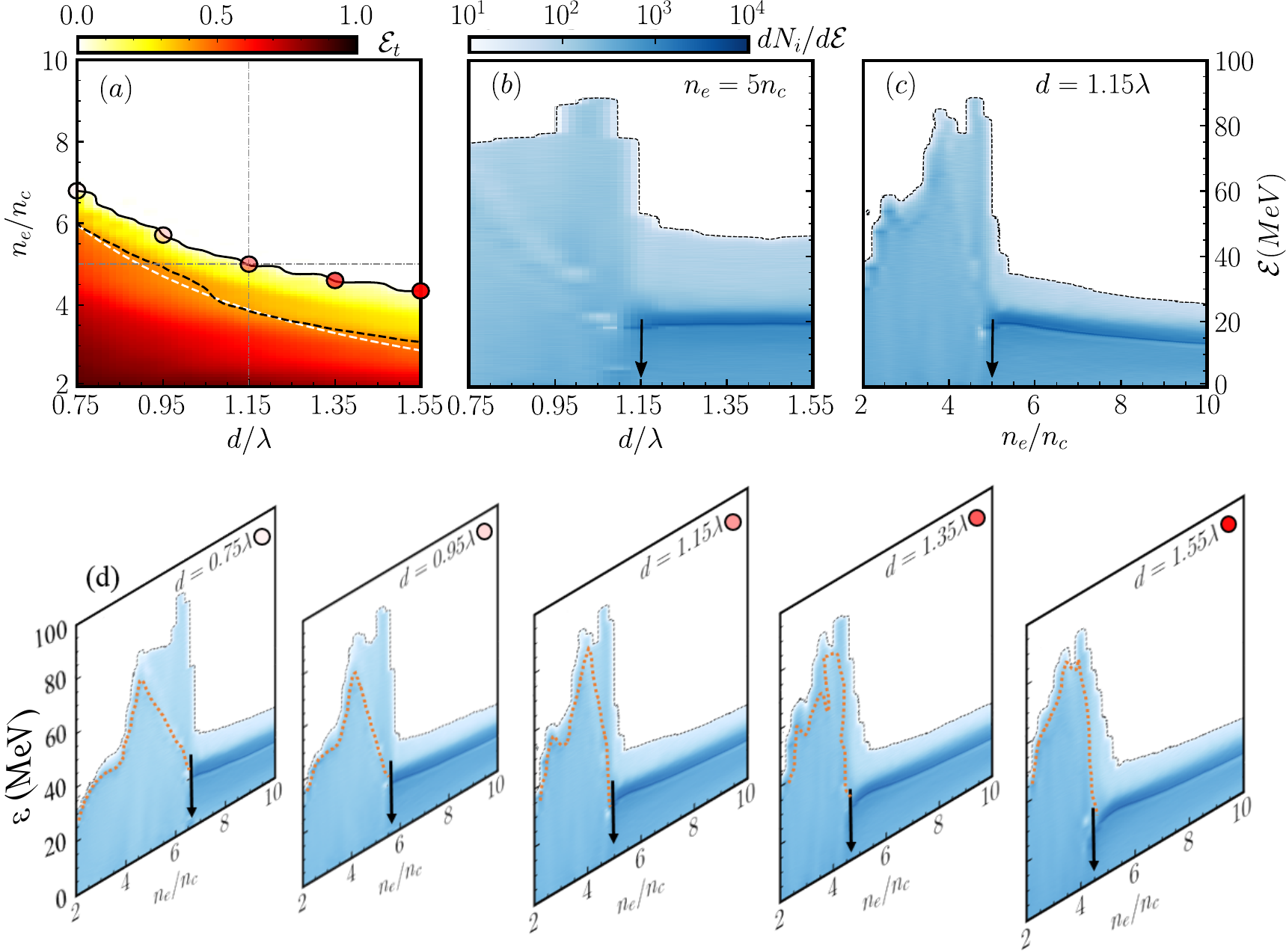}
    \vskip -0.3cm 
    \caption{Correlation between interaction regime and the nature of the resulting ion energy spectra: step density targets. (a) Transmitted laser energy fraction through the target with varying thickness (d/$\lambda$) and peak electron density ($n_e/n_c$) at simulation time 70$\tau$. The 2$\%$ (black solid line) and 40$\%$ (black dashed line) transmitted energy fractions are shown. The 40$\%$ black dashed line indicates threshold target density ($n_{th}$) for RIT for varying target thickness. Analytically predicted threshold target density (Eq. \ref{nth1}) is shown with a white dashed line and is consistent with the 40$\%$ iso-line obtained from PIC simulations. The colorbar shows the laser pulse energy transmission coefficient ($\mathcal{E}_t$) as in Eq. (\ref{tef}). The target density $n_e$ = 5$n_c$ and thickness  $d = 1.15\lambda$  are marked with horizontal and vertical grey lines in (a). In (b) ion energy spectral map with varying target thickness is presented for the threshold density of $n_e$ = 5$n_c$ (along the horizontal iso-density grey dashed line in (a)). In (c) ion energy spectral map for target thickness of $d = 1.15\lambda$ is presented with variation in target density (along the vertical iso-thickness grey dashed line in (a)). Note that the colormap for ion energy is plotted in logarithmic scale. The spectral map in (b) and (c) unequivocally shows that as the interaction enters from transparency into the opaque regime (across the  2$\%$ iso-transparency black solid curve in (a)) the accelerated ion energy spectra goes from exponential to quasi-monoenergetic peak structures irrespective of whether it is along the iso-density line or along the iso-thickness line. This establishes a consistent correlation between the regime of interaction and the nature of ion energy spectra over a wide range of parameter space. To probe this point further we plot the ion energy spectral map along different iso-thickness lines with varying peak target density. (d) shows the ion energy spectral map iso-thickness lines through the red semitransparent circles in the black curve separating the transparent and opaque regimes in (a). The black dotted line in (b), (c) and (d) represents the cut-off ion energy in both the regions, whereas the orange dashed contour lines on (d) marks the peak ion energy in relativistically transparent region. The black arrows in the ion energy spectral maps in (b), (c) and (d) demarcates the interaction conditions corresponding to the ($n_{e}/n_{c}, d/\lambda$) pairs on the black solid curve in (a) (identifies with target conditions indicated by each circle shown in (a)) emphasizing the transition phase between the regimes, i.e. from transparent to opaque conditions through the 2$\%$ iso-transmission curve. A clear correlation can be seen between $\mathcal{E}_t$ and. The colorbar shown above (d) illustrates the number of ions accelerated and represents for (b), (c) and (d). All the spectra are obtained at time, 70$\tau$.\label{STF}}
\end{figure*}
Now we look at the nature of ion energy spectra for the ions moving in direction of laser incidence. The ion energy spectra at a time when the peak of the laser pulse interacts with target (t=48$\tau$) are presented in black in Figure \ref{phasespace}(e) and (f), for both the cases (transparency and opaque). The ion energy first peaks and then follows a sharp cut-off, which is a typical spectral behavior of Hole-Boring RPA \cite{Weng_2012}. Since, up to the time t=48$\tau$ (peak interaction), the interaction is not yet over and transparency is not achieved, both the target conditions are in an overdense region, where we observed the spectral signature of the opaque region. At this intermediate interaction time, the ion cut-off energy is higher in the opaque region (Figure \ref{phasespace}(f)), than in the transparent case (Figure \ref{phasespace}(e)) due to the higher accelerating field operating at that instant. Also,  during the post-interaction phase (70$\tau$), in the transparency region (Figure \ref{phasespace}(e)), the accelerated ions follow a Plataea structure with peak energy of $\sim30$ MeV, whereas quasi-monoenergetic behavior in opaque region (Figure \ref{phasespace}(f)) with the peak ion energy of $\sim$16$\,$MeV. The peak ion energy from the transparent region and the cut-off ion energy from the opaque region is around $\sim$30$\,$MeV, in these particular cases. One should take into consideration that the respective ion energy spectra presented here correspond to spatially integrated velocity distributions at the relevant moments. Thus, in this section we have discussed the typical features of the interaction, the ion and electron density distributions in the real space-time domain, the relevant accelerating fields during and after the interaction and their consequence on the ion phase space distributions and resulting ion energy distributions for two special cases of interest. This has helped us identify the signatures of two different regimes of interaction. In the following we undertake a more systematic study of the effects of different target and laser parameters on the ion acceleration process, in the regimes of our interest trying to decipher the generic features that would allow us to design and establish an approach relevant for experiments.

\section{Influence of foil thickness and peak plasma density: the transition between the acceleration regimes}
\label{sec:steptarget}

In this sub-section, we investigate the effect of thickness variation in foil target on the ion acceleration mechanisms, namely  RIT and RPA. In ref. \cite{Cattani2000_PRE}, the stationary solutions for the 1D scenario are derived assuming the validity of a cold fluid model. Such a model operates under the approximations that, a CP monochromatic laser pulse interacts with the overdense plasma having a step-like electron density profile in a background of immobile ions. In the context of the cold fluid model, at this point, we would like to define a threshold electron density of the target $(n_{th})$. The threshold density is the maximum electron density that allows the laser pulse to transmit through the plasma target. The scaling law for the threshold density is given in ref. \cite{Siminos2012_PRE} in the case of a relativistic intense laser pulse ($a_0 >>$1) incident on semi-infinite plasma slab. A modified expression for threshold density incorporating the effect of the target thickness has been presented previously in a phenomenological way in \cite{Choudhary2016_EPJD} as:
\be n_{th} \sim \frac{2\lambda}{9d} \bigg(3 + \sqrt{9\sqrt{6} a_0 -12}\bigg) n_c \label{nth1} \ee
In the following we would use this expression as a model reference and we would also check the validity of the expression during our study, through a comparison using a large number of PIC simulations that we conduct. In this context, as a first step, we need to define clearly what we mean by transparent and opaque regimes over our parameter space of interaction.

In order to define the transparent and opaque regimes in a more quantitative and consistent way we define a parameter $\mathcal{E}_t$, called the transmitted energy fraction. We define the transmitted energy fraction as, 
\be \mathcal{E}_t = \frac{\int_{\eta_i}^{\eta_f} (E_y^2 + E_z^2) \mid_{TF} d\eta }{\int_{\eta_0}^{\eta_i} (E_y^2 + E_z^2) \mid_{IF} d\eta} \label{tef}\ee
where, $E_y$ and $E_z$ are the transverse components of the propagating laser fields, $\eta_0$, $\eta_i$ and $\eta_f$ are space-time representation for before (interaction has not started yet), beginning (initiation of the interaction) and end (interaction is over) of the interaction regions respectively, here, $\eta = t - x$ ($t$ and $x$ are in units of $\tau$ and $\lambda$ respectively implying $c=1$), the suffixes $IF$ and $TF$ correspond to the incident and transmitted fields respectively. In simulation time, $\eta_0$ = 0 $\tau$, $\eta_i$ = 36$\tau$, and $\eta_f$ = 60$\tau$. In Eq. \ref{tef}, the numerator $\int_{\eta_i}^{\eta_f} (E_y^2 + E_z^2) \mid_{TF}$ represents the transmitted fluence after the interaction with the target is over and the denominator $\int_{\eta_0}^{\eta_i} (E_y^2 + E_z^2) \mid_{IF} d\eta$ indicates the incident laser fluence on target, both expressed in the same units. Thus, here we have defined the transmitted energy fraction as the ratio between the transmitted laser energy and the incident laser energy. This parameter is directly calculated by post processing the PIC simulation results. For a given target and laser parameter, $\mathcal{E}_t=1$ implies total transmission and $\mathcal{E}_t=0$ indicates no transmission at all implying that laser energy is either totally reflected or absorbed. In a real interaction the value of the parameter lies between these two extremes, i.e. $0 \leq \mathcal{E}_t\leq1$. Thus, using the criterion of the transmitted energy fraction $\mathcal{E}_t$, of the laser pulse, we can differentiate between the region of transparency and opacity regimes in a quantitative manner.

The colormap in Figure \ref{STF}(a) shows the variation of the transmitted energy fraction, $\mathcal{E}_t$ when the interaction spans over a range of target peak electron densities, $n_e/n_c\in [2,10]$ and a sequence of target thicknesses, $d/\lambda \in [0.75,1.55]$ at each peak electron density. To begin with, each target peak plasma electron density and target thickness combination corresponds to a nominally overdense regime implying complete light reflection, in the case of a non-relativistic laser peak intensity, from the plasma critical density layer $n_{c}$. At our laser peak intensity relativistic effects come into play. For the given laser parameters, interaction on this two dimensional (2D) target parameter space clearly brings out several features. Firstly, along any vertical (increasing target peak electron densities along a line of constant target thickness) or horizontal line (increasing target thicknesses along a line of constant peak electron density) on the colormap in Figure \ref{STF}(a), the target becomes more reflective. Secondly, there is clearly demarcated opaque region on the 2D parameter space along with a gradually increasing transparent regime. In order to define transparency threshold we have defined an iso-line at, $\mathcal{E}_t=0.02$ below which we consider the target to be transparent. Thus, the black solid iso-line marked at 2$\%$ of transmitted energy fraction, indicates the maximum limit in threshold density ($n_{th}$) for the laser pulse to undergo transmission or reflection from the target. The 40$\%$ black dashed iso-line of transmitted energy fraction shows remarkable matching with the variation of threshold density obtained under the cold fluid approximation (the white dashed curve) using Eq. \ref{nth1}. Figure \ref{STF}(a) shows increasing target thickness ($d/\lambda$) approximately by a factor of two, the threshold density ($n_{th}$) reduces to nearly half, i.e. from 6$n_c$ at 0.75$\lambda$ to 3$n_c$ at 1.55$\lambda$. With increasing target thickness (0.75$\lambda$ - 1.55$\lambda$) and density (2$n_c$ - 10$n_c$), increases the target areal density ($\propto n_e d$). Therefore, the incident laser field on the target is insufficient to remove the substantial amount of electrons from the target to achieve RIT and thereby reflects the pulse from the surface of the thicker target with higher electron density. This regime of ion acceleration lies under hole-boring (HB) RPA. Thus, one can effectively control the transition from transparency to opacity by tuning the target thickness and density.

We now investigate the correlation between interaction regime and the nature of the ion energy spectra, in order to see whether the understanding developed in the previous section for two specific cases, can be validated over a wider parameter range. To elucidate this process of transition, we present the ion energy spectral map in Figure \ref{STF}(b) at a fixed electron density($n_e$  = 5$n_c$) and varying target thickness ($d/\lambda$). The target density, in this case, is chosen in such a way that it can cover both the regions, i.e., transparent (RIT) and opaque (RPA), over the range of target thickness (along the horizontal grey dashed line in Figure \ref{STF}(a)). For the first half of thickness variation \textit{i.e.,} for $d <$ 1.15$\lambda$, the condition satisfies RIT where the cutoff energy reaches maximum up to 90$\,$MeV following the energy distribution as previously observed in Figure \ref{phasespace}(e). On the other hand, for the target thickness of $d >$ 1.15$\lambda$ a quasi-monoenergetic spectrum is observed akin to that in Figure \ref{phasespace}(f), with the constant peak ion energy peak at 20$\,$MeV and cut-off energy of $\sim 55\,$MeV. It is evident from Figure \ref{STF}(a) as well, that for $n_e = 5n_c$ the , the iso-density line crosses the transparency threshold curve ($\mathcal{E}_t=0.02$) at d = 1.15$\lambda$, target start to become opaque for all the thicknesses $d> 1.15 \lambda$. Therefore, we can refer to thickness d = 1.15$\lambda$, as a transitioning thickness between these two regimes. This point representing the change in the ion spectra is marked with an arrow in \ref{STF}(b).

Similarly, Figure \ref{STF}(c) shows the ion energy spectral map for fixed target thickness (d = 1.15$\lambda$) while increasing the target density from 2$n_c$ to 10$n_c$. We observe similar behavior in the ion energy spectrum as in the case of fixed density (5$n_c$) target shown in Figure \ref{STF}(b). Since $n_e = 5n_c$ is the threshold density (marked with an arrow) for target thickness of 1.15$\lambda$, it acts as a transitioning density point. For $n_e <5n_c$ the target undergoes transparency, and we observe the representative ion energy spectrum with cut-off energy increasing from 40 MeV for near critical density target to 90 MeV for relatively higher density target. Whereas, in case of $n_e > 5n_c$ spectral features highlight quasi-monoenergetic behavior. However, in this case, we observe the reduced cutoff energy from 40 MeV (threshold density $n_e$ = 5$n_c$) to 25 MeV (overdense target $n_e = 10n_c$) as well as peak ion energy from 20 MeV to 15 MeV. Since, the target densities $n_e >5n_c$ lies under opaque region, the laser ponderomotive push is not strong enough to remove all the electrons from the target. This results in weak charge separation field which consequently restricts the ion energies within certain limit.

In addition, we presented the ion energy spectral map with varying target density for four different target thicknesses in Figure \ref{STF}(d) corresponding to the thicknesses indicated with circles filled in different shades of red, lying on the threshold density curve in Figure \ref{STF}(a). These exhibits the typical nature of energy distribution in transparent region (RIT) and quasi-monoenergetic distribution of ion energy spectrum in reflected regime (RPA). For each target thickness, the threshold density acts as a transitioning point, which is consistent with the transmitted energy fraction iso-line (2$\%$) in Figure \ref{STF}(a). Additionally, we observed that the peak ion energy (yellow dashed in transparent region and dark blue in opaque region) follows the similar trend as of ion cut-off energies (black dotted lines) for all the target thicknesses, over the varying target densities ($2n_c - 10n_c$ ). 

Therefore, the maximum ion cut-off energy at the target threshold densities for all the target thicknesses (0.75$\lambda$, 0.95$\lambda$, 1.15$\lambda$, 1.35$\lambda$, and 1.55$\lambda$), reaches up to 90  MeV. Additionally, the maximum ion cut-off energy was obtained in RIT domain, ranging from $\sim$40  MeV to $\sim$90  MeV, which is significantly higher than that achieved in HB-RPA ($\sim$25  MeV to $\sim$40  MeV). In other words, the transition point between RIT and RPA regime is the key criteria for achieving maximum ion energy, where RIT determines the maximum ion cut-off energy and HB-RPA region is the criteria for achieving quasi-monoenergetic peak ion energy (maximum $\sim$20 MeV). Hence, one can benefit from the transition between different regimes of ion acceleration, depending upon the ion energy requirement for the given the experimental conditions. 

\section{Controlling proton acceleration: foil with plasma density gradient}
\label{sec:gradient}

 \begin{figure*}[t!]
    \vskip -0.35cm
    \centering
        \includegraphics[scale=0.75]{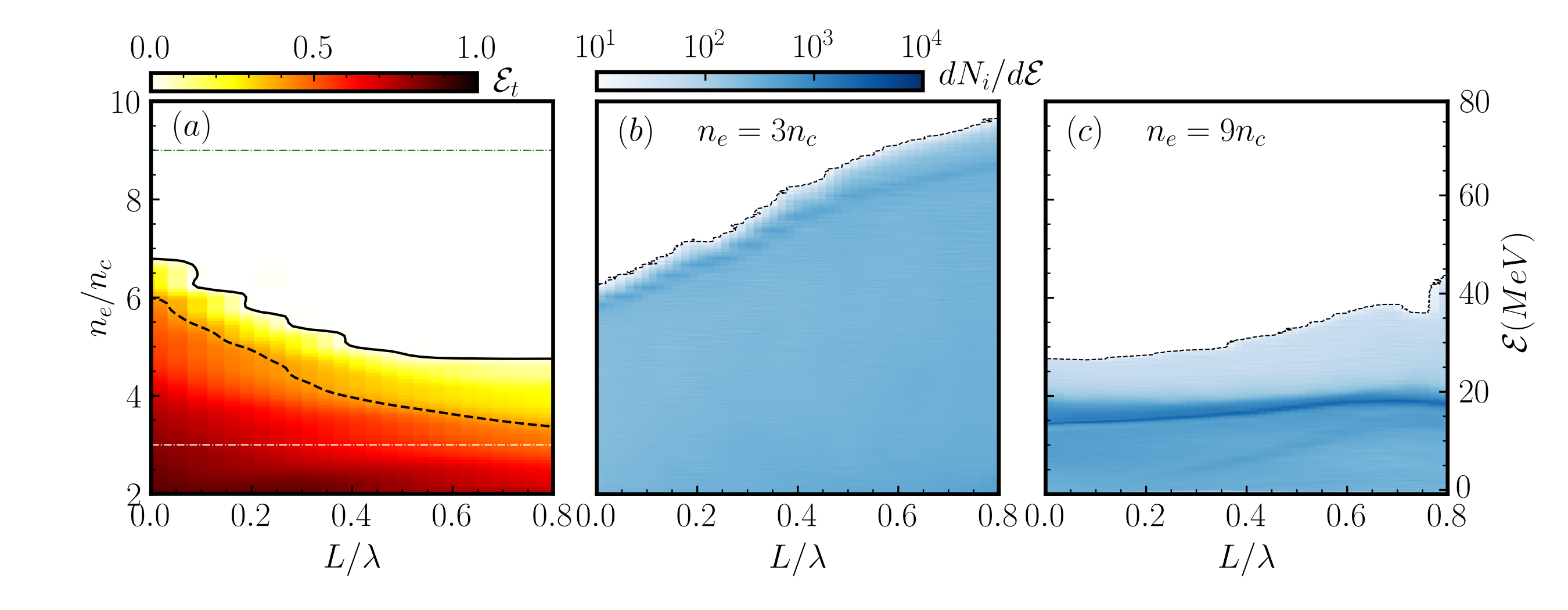}
    \vskip -0.35cm 
    \caption{Correlation between interaction regime and the nature of the resulting ion energy spectra: foils with varying plasma gradient scale lengths. (a) Transmitted laser energy fraction for the case of target varying scale-length (L/$\lambda$) and density ($n_e/n_c$). The colorbar is the energy transmission coefficient ($\mathcal{E}_t$) as defined in Eq. \ref{tef}. The black solid line corresponds to the 2$\%$ of transmitted energy fraction, and the black dashed line is for the 40$\%$ transmitted energy fraction. (b) and (c) show the ion energy spectral maps with varying scale-lengths for $n_e$ = 3$n_c$ (dashed white line (a)) and 9$n_c$ (dashed green line (a)) respectively. The black dashed contour lines in (b) and(c) demarcate the maximum or cut-off ion energy in each case. The colorbar on top of Figure (b) represents the number of particles accelerated per energy bin at the corresponding energy.\label{TF}}
\end{figure*}
Under a realistic experimental scenario, the driving laser pulse always have a limited temporal contrast, which leads to the expansion of the target front surface and giving rise to the plasma gradient before the peak of the main laser is incident on the target. In case of thin targets that are essential for applications relevant for ion acceleration with mechanisms such as hole-boring \cite{pukhov1997laser} or relativistic transparency\cite{fernandez2017laser}, the target can be destroyed before the peak of the laser pulse can interact, if the laser contrast is poor. Influence of pulse temporal contrast on the acceleration process in the case of optically thicker targets has been observed experimentally \cite{McKenna2006}. Thus, in all these experiments a high temporal contrast of the main interacting laser, and a separate fine control of the target plasma density gradient \cite{Kahaly2013} are the prerequisites. Thus the impact of plasma density gradient on the physics under discussion cannot be overemphasized. In this section we look into the influence of plasma density gradient on ion acceleration process.

In this study, we use a $0.75\lambda$ thick foil target with an exponential density profile, $n(x) = n_{0} exp(-(x-35)/L)$, at the front side of the target as an extra controlling parameter for the interaction. Where $n_{0}$ and $L$ are the peak electron density and plasma scale-length, respectively. In Figure \ref{TF}(a), we vary the target peak electron density and the plasma density scale-length ($L/\lambda$) and plot the transmitted energy fraction as a function of these two parameters. Altering the target density gradient, significantly affects the threshold density criteria. Similar to the case of thickness variation for the step target density profile, here as well we sketch an iso-line along the different scale-lengths for 2$\%$ and 40 $\%$ of light energy transmission through the target. With increasing scale-length, the transmitted energy fraction of the laser pulse decreases, for a fixed laser intensity. Subsequently, the target threshold density ($n_{th}$) reduces by approximately 30$\%$, due to the increase in overall areal density of the target. Also, we observe that the iso-lines (2$\%$ and 40$\%$) for the case of varying scale-length (Figure \ref{TF}(a)) showed slightly higher threshold density while compared with the step-like target thickness variation case (Figure \ref{STF}(a)).

Further, Figure \ref{TF}(b) and (c) exhibits the spectral features of ion energy as a function of scale-length for relativistically transparent ($n_e = 3n_c$ marked as the white dashed line in Figure \ref{TF}(a)) (b) and overdense ($n_e = 9n_c$ marked as the green dashed line in Figure \ref{TF}(a)) (c) regions respectively. In Figure \ref{TF}(b), the ion energy spectra show features similar to that in Figure \ref{phasespace}(e). In this case, the ion cut-off energy increases almost linearly from $\sim$45 MeV to $\sim$80  MeV with increasing scale-length. The peak ion energy increases from $\sim$40 MeV to $\sim$65 MeV, closely following the behaviour of cut-off energy. On the other hand, in Figure \ref{TF}(c) the ion energy spectra, in the overdense region ($n_e$ = 9$n_c$), shows significantly low increase in the ion cut-off energy from $\sim$30 MeV to $\sim$40 MeV over the entire range of target scale-length. Although, the peak ion energy is maintained nearly at $\sim$ 19 MeV with energy spread $(\Delta \mathcal{E} / \mathcal{E})\%$ = 2.8$\%$ for sharp gradient(L$/\lambda$ = 0) and 7.4$\%$ for long gradient ($L/\lambda = 0.80$), where $\mathcal{E}$ is the peak ion energy and $\Delta \mathcal{E}$ is the FWHM of the peak energy. This highlights the quasi-monoenergetic feature from the overdense plasma target. In addition, for a fixed scale-length, if we scan the target densities (2$n_c$ - 10$n_c$), one can transit from transparency to opacity region interchangeably in a continuous manner. Thus, changing the ion energy spectra to quasi-monoenergetic distribution. Therefore, control of $L/\lambda$, which can be achieved experimentally very easily by controlling the delay of a prepulse arriving before the main pulse starts interacting with the target, acts as an extra optimization parameter for the production of energetic protons. 

\section{Chirp control of foil dynamics on Gradient Target}
\label{sec:chirp_main}
\begin{figure*}[t!]
	\vskip -0.1cm
	\centering
    	\includegraphics[scale=0.90]{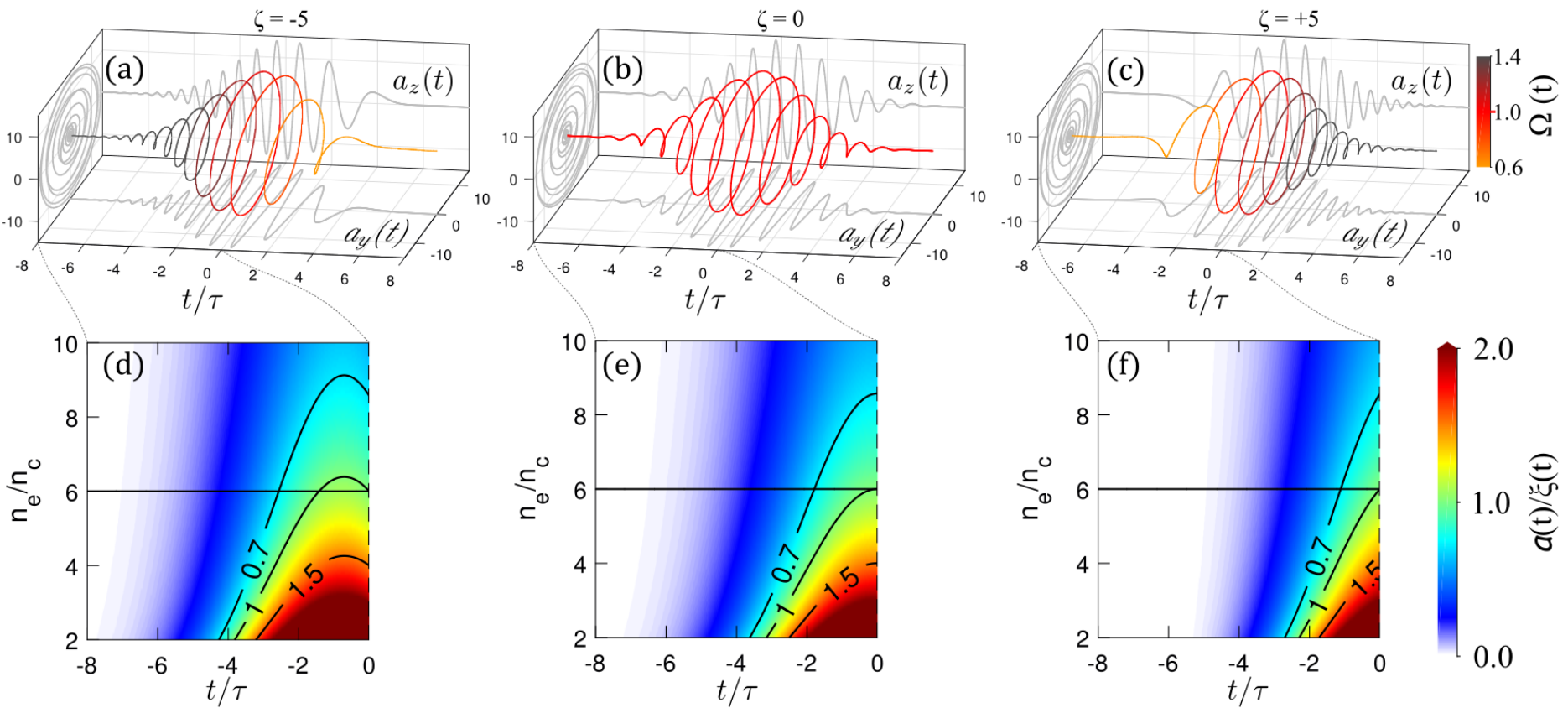}
	\vskip -0.3cm 
	\caption{Simplified picture elucidating the influence of laser chirp. Upper panel: The temporal profile of a circularly polarized (a) negatively chirped ($\zeta=-5$), (b) unchirped ($\zeta=0$), and (c) positively chirped ($\zeta=+5$) laser pulse are shown. The projected transverse field components $a_{y}(t)$ and $a_{z}(t)$ are presented in gray solid lines, plotted on the respective transverse planes. The variation in time dependent frequency of the laser pulse as shown in color bar ($\Omega(t)$ as defined in Eq. \ref{freq}) is encoded on the color variation across the circularly polarised laser field. The pulse peak here is located at $t=0$. $\Omega(t)=1$ represents the carrier frequency and -ve $t$ indicates early in the interaction. Lower panel: The colormaps plot the ratio of the time dependent laser pulse amplitude $a(t)$ and the parameter $\xi(t)$ (defined in Eq.\ref{k1}) during the first half of the interaction (the time axes are zoomed into the range from the start of the laser pulse at -8$\tau$  up to the peak of the pulse intensity at 0$\tau$) for the step target case over a range of target densities for (d) negatively chirped, (e) unchirped, and (f) positively chirped pulse scenarios. The black contour lines are plotted at $a(t)/\xi(t)=$ 0.7 (in opaque region), 1.0  (in threshold region), and 1.5 (in transparent region) for the fixed density case of $n_{e}=6n_{c}$ (black horizontal solid line). As is clearly seen, if target remains unchanged during the interaction, in (d) for the negatively chirped case, before the peak of the laser pulse intensity, the interaction at an initial target density of $n_{e}=6n_{c}$ enters the transparent regime of operation. (e) The transparent regime is approached near the peak of the pulse for $\zeta=0$ and (f) the same target meets the transparency condition after the peak of the laser pulse for $\zeta=+5$.\label{pulse}}
\end{figure*}

In addition, to the control over target plasma characteristics, ion acceleration in laser-plasma interaction can greatly be optimized by tuning the drive laser optical parameters  \cite{Galow2011_PRL, Vosoughian2015_POP,Mackenroth2016_PRL, Choudhary2018_POP}. The influence of laser polarization, pulse duration and peak intensity under different target conditions have been touched upon previously \cite{Liseikina2007, macchi_ls,Macchi201041,Sahai2013_PRE,Culfa2021,Almassarani2021} and the potential of the relativistic transparency regime have been demonstrated experimentally \cite{Gonzalez2016_nature,Higginson2018_nature}. Here, we introduce a frequency chirp in the driver pulse, which is relatively simpler to control in the experiment and investigate its efect on the acceleration process. The frequency chirp is introduced, within the phase function $\phi(\eta) = 2\pi [\eta+g(\eta,\zeta)]$ in the expression of the laser field in Eq. \ref{cf}, where, the chirp function $g(\eta,\zeta)$, is defined in Eq. \ref{cf} below:
\begin{equation}
\begin{aligned}
   g(\eta,\zeta) = \bigg(\zeta \bigg[4 ln (2) \frac{\eta^{2}}{\tau_{FWHM}^2} &+ \frac{\pi^{2}\tau_{FWHM}^2}{4 ln (2) (1 + \zeta^2)}\bigg] \\
   & + \frac{tan^{-1}(\zeta)}{2}\bigg)\frac{1}{2 \pi}
\label{cf}
\end{aligned}
\end{equation}
where, $\zeta$ is the chirp parameter and $\Omega(\eta)$ is the instantaneous frequency given by,
\begin{equation}
\begin{aligned} 
\Omega(\eta) = \frac{1}{2 \pi} \frac{\partial \phi(\eta)}{\partial \eta} &= 1+\frac{\partial g(\eta,\zeta)}{\partial \eta} \\
&= 1 + \zeta\frac{4 ln (2)}{\pi \tau_{FWHM}^2}\eta 
\label{freq}
\end{aligned}
\end{equation}

In these units, $\Omega = 1$ represents the central (at the peak of the pulse) carrier frequency of the pulse in the unchirped ($\zeta=0$) case. This functional definition was first proposed in ref.\cite{Mackenroth2016_PRL} as a chirped plane wave model. However, a modification is made to take advantage of preserving peak field amplitude of the pulse during any variation in the chirp \cite{Choudhary2018_POP}. This is done in order to ensure that when we investigate the effect of chirp parameter $\zeta$ on the interaction, we can keep the other laser parameters fixed. For simplifying the representation in the following discussion we interchange $\eta$ with $t$ in the expressions without any loss of generality. 

With the chirped Gaussian pulse, defined above, we have modified the PIC code to incorporate such a CP field and eventually perform a series of simulations to investigate the effect of frequency chirp on ion acceleration. In Figure \ref{pulse}, we have presented the (a) negatively chirped, (b) unchirped and (c) positively chirped laser pulse, that have been used in the simulation. The time dependent instantaneous frequency of the pulse ($\Omega(t)$) is shown with the color axis Figure \ref{pulse} (a-c). In these particular plots the peak of the laser field envelop is located at $t=0$ and the negative value of time axis indicates `earlier' in the interaction. The chirp is defined in such a way that for a negatively (positively) chirped pulse, initially high (low) frequency component interacts with the target followed by the low (high) frequency component.  

Before we delve into the results of fully relativistic PIC simulations, we discuss here pedagogically the impact chirp might have by extending the constant parameter $\xi_{0}$, introduced earlier, to an equivalent time dependent form. Since, a chirp in the laser pulse signifies the change of laser frequency with time, the corresponding plasma critical density also becomes time dependent, $n_{c}(t)=(n_{c})_{\omega_{0}}\Omega(t)^{2}$. The chirped CP plane wave pulse can be conceptually considered as a superposition of all the monochromatic CP plane waves constructed with strengths proportional to the instantaneous field envelops and frequencies equal to the instantaneous frequencies at all times. Hence, the constant parameter $\xi_0$, also changes and is redefined to its instantaneous form as follows,
\be
\xi(t) = \bigg(\frac{n_e}{n_c}\bigg)_{\Omega(t)} \frac{d\pi}{\lambda_{\Omega(t)}} = \bigg(\frac{n_e}{n_c}\bigg)_{\omega_{0}} \frac{d\pi}{\lambda}\frac{1}{\Omega(t)} = \frac{\xi_{0}}{\Omega(t)}
\label{k1}
\ee
Thus the ratio relevant to the condition of transparency discussed previously in section \ref{sec:foil} now becomes, $\frac{a(t)}{\xi(t)} = \frac{ a(t) \Omega(t)} {\xi_0}$.
The colormaps in Figure \ref{pulse} plot the value of this ratio for a step target with initial thickness of $d=0.75\lambda$ and different target peak electron densities at different instants of time. We emphasize here, that this oversimplified description does not take into account, among many things, the dynamics of the target and takes each time instant it treats the situation as stationary. Nonetheless, as we would see that the discussion provides us some insight on the effect of the sign of chirp on the interaction. The different cases for negatively chirped ($\zeta = -5$), unchirped ($\zeta = 0$) and positively chirped ($\zeta = +5$) pulse under the condition of monochromatic pulse and step-like target of finite thickness, are depicted in Figure \ref{pulse}(d), (e) and (f) respectively.

The main difference between transparency and opaque regime of operation can be distinguished by the ratio of time dependent laser pulse amplitude $a(t)$ and the parameter $\xi(t)$. 
As mentioned in ref. \cite{macchi_ls}, the condition $a_0/ \xi_0 > 1$, indicates the region of operation in relativistic self-induced transparency and $a_0/ \xi_0 \leq 1$ refers to radiation pressure acceleration, respectively. Therefore, in Figure \ref{pulse}(d-f), we show the time dependent ratio of $a(t)$ to $\xi(t)$ in colormap for varying target density ($n_e/n_c$) and time ($t/\tau$) in order to identify the instant of time, if any until the peak of the pulse, at which the target reaches the condition of transparency. In our present illustration, the pulse reaches peak amplitude at $t = 0$. We consider the time up to the peak amplitude, i.e. from $-8\tau$ to $0\tau$ to plot the ratio of $a(t)$/$\xi(t)$. Now, to understand the region of transparency and opacity, we specifically consider the case for a peak density of $6n_c$, shown with a solid horizontal black line along with three important regions marked with contour lines at $a(t)/ \xi(t) = 0.7$ in opaque region, $a(t)/ \xi(t) = 1.0$, transition or threshold region and $a(t)/ \xi(t) = 1.5$, transparent region in Figure \ref{pulse}(d-f). 

Before proceeding any further, we first make several observations. Firstly, in case of negatively chirped pulse, Figure \ref{pulse}(d), since higher frequency components of the pulse is interacting at the beginning ($t \leq 0$), therefore the majority of field cycles interacts up to $t = 0$.  Whereas, for the positively chirped pulse Figure \ref{pulse}(f) low-frequency cycles of pulse interacts with the target initially, therefore the number of field cycles interacted up to the peak  $t \leq 0$ is much less as compared to \ref{pulse}(d). Secondly, according to the ratio ($a(t)/ \xi(t)$), at the peak field ($t=0\tau$) instance for unchirped pulse (Figure \ref{pulse}(e)) at $n_e=6n_c$ (horizontal black line) the transparency condition is achieved.  Whereas, for negatively (Figure \ref{pulse}(d) and positively chirped pulse (Figure \ref{pulse}(f), the transparency condition is achieved at slightly higher target density (approximately at $n_e$=6.25$n_c$). Thus, we note that this crossover is before $t = 0\tau$ for negatively chirped pulse and after $t = 0 \tau$ in positively chirped pulse. 
\begin{figure}[t!]
	\vskip -0.35cm 
	\centering
      \includegraphics[scale=0.58]{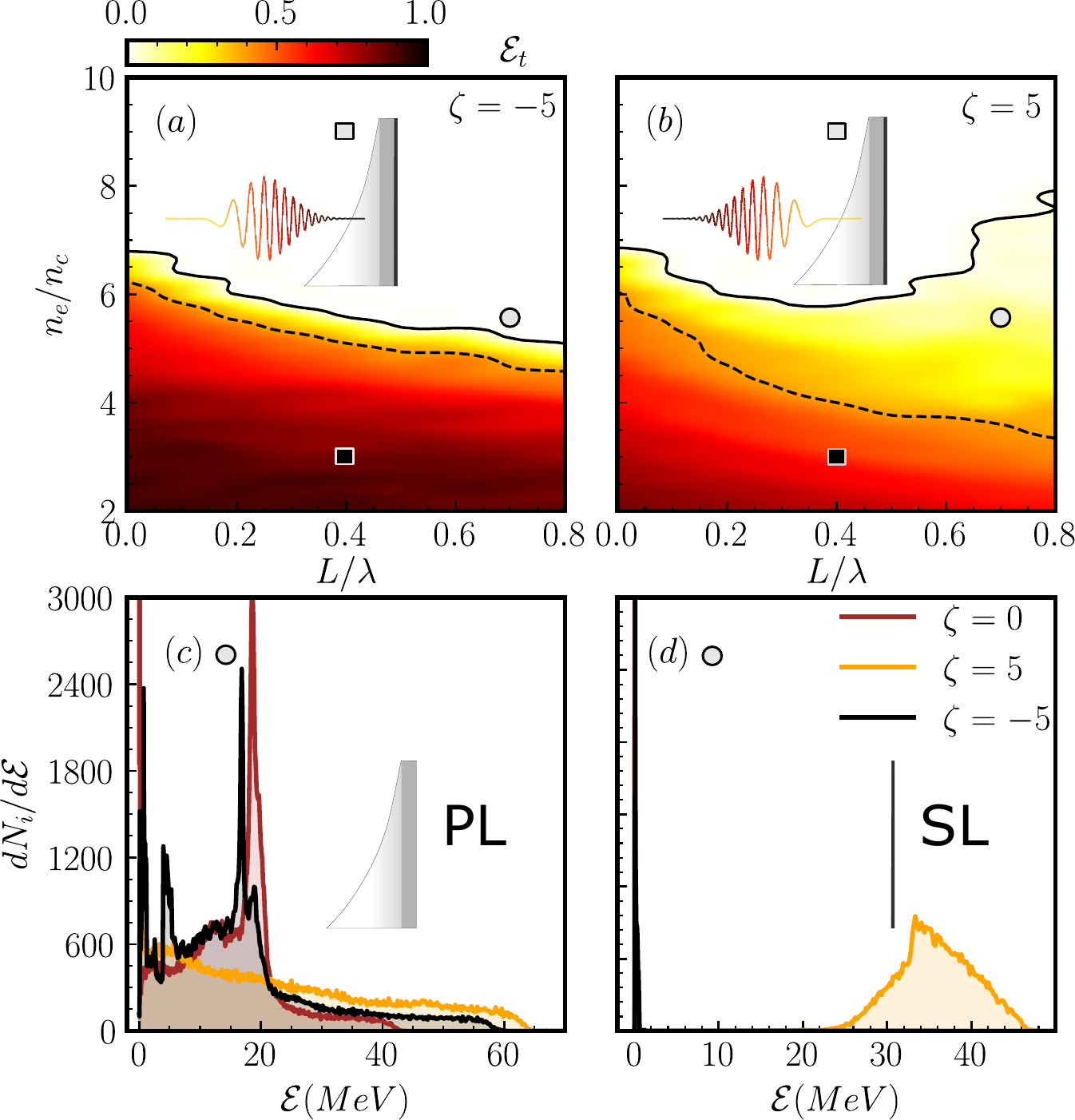}
   	\vskip -0.3cm 
   \caption{Laser pulse transmitted energy fraction ($\mathcal{E}(t)$) for (a) negatively and (b) positively chirped pulse with varying target density ($n_e/n_c$) and scale-length (L/$\lambda$). Similar to the above figures, the iso-lines for transmitted laser energy fraction are marked at 2 $\%$ (black solid) and 40 $\%$(black dashed). The shaded squares (black square with grey fill or the reverse) on the colormap in (a) and (b) indicate the specific transparency conditions corresponding to the primary target electron density. The point marked in black solid circle with grey fill in (a) and (b) identifies the target condition where the transparency difference due to positive and negative chirp is maximum. Ion energy spectrum from (c) the primary layer (PL) (blue square) having peak density = $5.6n_c$ and scale-length = $0.7\lambda$ and, (d) secondary layer (SL) of thickness = $0.2\lambda$ and density = $0.1n_c$, for unchirped (maroon solid), positively chirped (yellow solid) and negatively chirped (black solid) pulses. The spectra are obtained at 70$\tau$, simulation time.\label{chirp-ene}}
     \vskip -0.55cm
\end{figure}
In the real interaction, for the normally incident CP laser pulse, the laser field would dynamically propagate into the target plasma upto the critical electron density layer before getting reflected. From the surface of reflection the light field would penetrate upto skin depth and the ponderomotive push of the laser would pile up electrons inside the target leading to a dynamic in the electron density profile \cite{Macchi2009CR,Vincenti2014_Nat}. For $\zeta = -5$, the higher frequency initial part would only start piling up electrons after propagating deeper inside the target compared to the lower frequency trailing part of the laser. On the other hand at, $\zeta = +5$ the lower frequency leading part would start spiking the electron density, which when optimally done, would reflect the following higher frequency light more efficiently. Thus, unlike the simplistic picture presented in Figure \ref{pulse}(d-f), in case of a longer ramp in target electron density, the laser ponderomotive pressure can be expected to be more effective in the case of $\zeta = +5$ in pushing the electrons out of the target and rendering the target transparent, compared to the case when  $\zeta = -5$, since a longer ramp helps the snow ploughing effect \cite{Kahaly2013,Vincenti2014_Nat}. Thus, a priory one might expect lesser dependence on chirp in case of step density targets. In the following, we would study this remarkable effect of laser chirp on the previously mentioned step-like foil target having a front density gradient akin to section \ref{sec:gradient}. 

In Figure \ref{STF}(a) and Figure \ref{TF}(a) we have already discussed the effect of target thickness and density gradient on the transmitted energy fraction, in the case of unchirped CP pulses. Now, we expand this idea to the chirped laser pulses. In Figure \ref{chirp-ene}(a) the transmitted pulse energy fraction for negatively chirped pulse and (b) for positively chirped pulse over variable scale lengths and densities are presented. In both the cases, two iso-lines are drawn at 2$\%$ (black solid) and 40$\%$ (black dashed) transmitted laser pulse energy. For a negatively chirped pulse, in Figure \ref{chirp-ene}(a), the threshold density reduces only about $10~\%$ over the entire range of scale-length. While comparing with Figure \ref{TF}(a), the changes in threshold density with scale-length is minimal. This is evident in transparent region, where we observe that up to $4n_c$ almost entire pulse energy is transmitted through the target for all the scale-lengths ($L = 0.0 \lambda$ to $0.8 \lambda$). On the other hand, for the case of positively chirped pulse, Figure \ref{chirp-ene}(b), at 40$\%$ laser energy transmission (black dashed), we see the reduction in threshold density of about 50$\%$ for varying target scale-length. Whereas, for 2$\%$ iso-line (black solid), we first observe the decrease of $\sim 14 \%$ in target threshold density up to $L = 0.4 \lambda$. Later, from $L = 0.4\lambda$ to $0.8\lambda$, the threshold density increases by nearly $33 \%$. Thereby, the gap between the 2 $\%$ and 40 $\%$ transmitted energy fraction iso-lines widens in comparison with the negatively chirped pulse.
This effect is due to continuous compression and pilling of electrons at the rear side from the low frequency cycles at the beginning of the interaction with the target, followed by the high frequency cycles which pushes them out of the target surface, thereby resulting in continuous transmission of laser energy from the target.

We have observed that, while considering the 2$ \%$ iso-line one can achieve is a higher value of threshold density in the positively chirped pulse case. This implies that experimentally it could be possible switch the domain of interaction from opaque to transparent regime just by controlling the chirp parameter. One such contrasting point in the laser-matter parameter space is indicated using the black circles with grey fill in Figures \ref{chirp-ene}(a-b) (representing peak density of 5.6$n_c$ and scale length as 0.7$\lambda$). In previous sections we have observed that, over the parameter space of interest to this work, the accelerated ions show quasi mono energetic spectral features in the opaque regime of operation. Whereas in the transparent regime the ion spectra have shown flat behaviour and sharp cutoffs. Nonetheless, we notice that the bipolar accelerating field generated in the transparent case as shown in Figure \ref{fieldens}(a), shows a nice smooth peak profile located at the exit end of the target ion density profile. This gives us the idea of adding a thin low density layer at the back of our initial target, and see whether we can benefit from this accelerating charge separation field and accelerate ions of choice. Hence to probe if the transmitted laser pulse can allow us to accelerate quasi mono-energetic ions we add a thin layer (thickness 0.2$\lambda$) of low-density ($n_e=0.1n_c$) Hydrogen behind the main target of thickness 0.75$\lambda$. For the main target we use variable peak densities (2$n_c$ to 10$n_c$) and gradient scale-lengths (0$\lambda$ to 0.8$\lambda$). This results in re-configuring of target geometry to the double layer target. The choice of the parameters of this thin layer are made in such a way that the electrostatic field created by the main target layer remains unaltered. In rest, we will refer to the main target as primary layer (PL) and additional target as secondary layer (SL). 

In Figure \ref{chirp-ene}(c,d) we present the ion energy spectra for the cases of unchirped ($\zeta=0$), positive ($\zeta=5$) and negative ($\zeta=-5$) chirped pulses. The respective ion energy spectra from the PL with peak density of 5.6$n_c$ and scale length as 0.7$\lambda$ is shown in Figure \ref{chirp-ene}(c). In Figure \ref{chirp-ene}(d) we show ion energy spectra from the SL, with above-mentioned parameters. Here, we choose the PL parameter in such a way that approximately 20$\%$ of the laser pulse energy propagates through the target (see the black circle with grey fill in Figure \ref{chirp-ene}(a-b). As mentioned previously, in this case, only positively chirped pulse allows the transmission of laser pulse through the PL. For the case of negatively chirped and unchirped pulse, the transmission coefficient is close to zero. Previously in section \ref{sec:foil} we have already established that, in the transmission region, the ion energy spectrum shows the near Maxwellian distribution, whereas a quasi mono-energetic feature is observed in the opaque region. In Figure \ref{chirp-ene}(c), we observe similar characteristics of Maxwellian distribution in ion energy spectrum for positively chirped pulse and quasi-monoenergetic features from negative and unchirped laser pulse. The cut-off ion energy is slightly higher than 60 MeV for positively chirped pulse and approximately 60 MeV for negatively chirped pulse. Whereas, for unchirped pulse, the ion cut-off energy is around 40 MeV. Thus, one can utilize chirped pulses to effectively enhance the ion cut-off energy in both the regimes (transparency and opacity).  

Furthermore, in Figure \ref{chirp-ene}(d) we show the significance of the SL in the transparency regime. Since, only positive chirp pulse allows the laser to transmit through the PL for a specific set of parameters as indicated in Figure \ref{chirp-ene}(c), therefore, a mono-energetic ion bunch with a peak energy at 35 MeV and cut-off energy reaching close to 45 MeV is observed. Thus, with a secondary layer along with a target of suitable parameters, one can obtain  mono-energetic bunches in a controlled manner. On the contrary, for the case of negatively chirped and unchirped pulse, since the PL parameters lies in the opaque region, we observe a clear quasi-monoenergetic spectral signature in ion spectra from PL peaking at 20 MeV (Figure \ref{chirp-ene}(c)) and negligible ion energies from the SL (Figure \ref{chirp-ene}(d)).
\begin{figure}[t!]
    \includegraphics[scale=0.65]{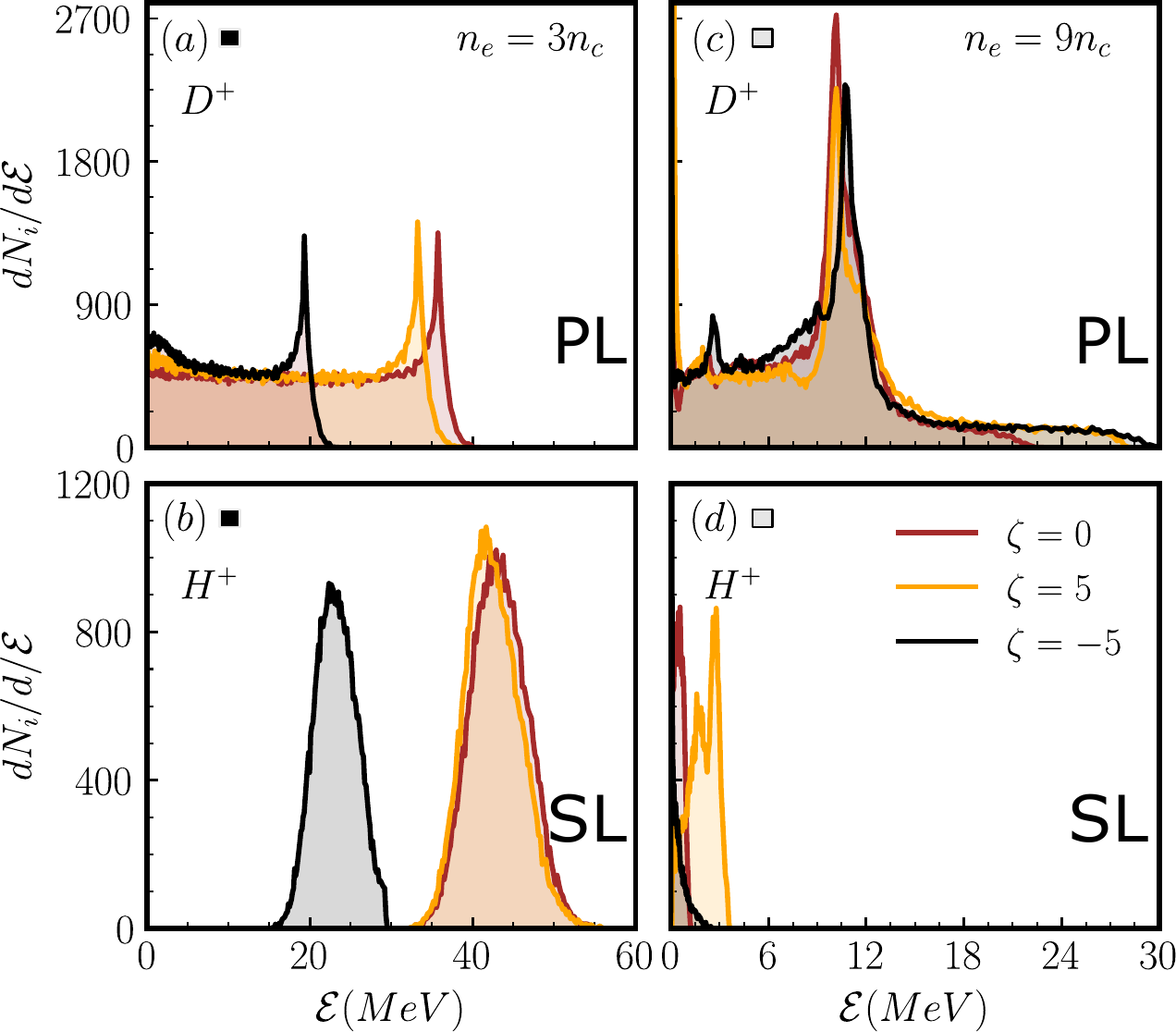}
    \vskip -0.3cm 
    \caption{The ion energy spectra from double layer target, composed of Deuterium (PL) and thin, low-density ($\mathrm{0.2\lambda}$, $0.1n_c$) Hydrogen (SL). (a,b) and (c,d) correspond to ion energy spectra from both the layers, for primary layers with peak electron density $3 n_c$, and $9n_c$ respectively, with scale-length of 0.4$\lambda$ in all the cases. The color of the traces indicate the laser chirp conditions: positively chirped (yellow solid), unchirped (maroon solid) and negatively chirped (black solid) cases. The shaded squares (black square with grey fill or the reverse) on the top left corner of each sub-figure correlates with those marked in Figure \ref{chirp-ene}(a,b) identifying the transparency conditions corresponding to the primary target electron density.\label{energy}}
     \vskip -0.5cm
\end{figure}

\section{Tuning into Quasi-Monoenergetic ion spectrum}
\label{sec:tuning}

The key aspect we observed in the section above is that, using chirped pulses one can precisely control the laser transmission through the target with specific properties or in other words, enhance the threshold density of the target for the incident laser pulse. In Figure \ref{chirp-ene}(d), we also observed that using an additional layer behind the main target (double layer target) leads to the generation of mono-energetic ion bunches in the transparency region. Detailed modelling on the optimization of a double layer target parameter in a different context has been proposed \cite{Andrea2020}. Here we keep the configuration of the second layer fixed and look more into the physics aspects. More specifically in order to gain further control over ion energy spectra, we investigate spectral features from both the layers independently while controlling the target species and establish the proof of principle.

Towards achieving this goal, we have used a target to be composed of two layers of different species, Deuterium in primary layer (PL) and Hydrogen in secondary layer (SL). The target peak densities of the PL are similar to previous section \ref{sec:gradient}, $n_e = 3n_c$, corresponding to transparent region and $n_e = 9n_c$, corresponding to opaque region, respectively. Also, the double layer thickness (PL : $0.75\lambda$, SL : $0.2\lambda$) is kept similar to Figure \ref{chirp-ene}(c,d) with scale-length of $L = 0.4\lambda$. The ion energy spectra are presented in Figure \ref{energy} for unchirped, positive and negatively chirped pulses, where (a,c) is for Deuterium target (PL) and (b,d) corresponds to Hydrogen layer (SL). In Figure \ref{energy}(a), we have observed the spectral signature similar to Maxwellian distribution, peaking near the cut-off energy as in Figure \ref{phasespace}(e) at 70$\tau$. For the case of negatively chirped pulse, 50$\%$ reduction in ion cut-off energy was observed while comparing with positively chirped and unchirped pulse. Since, for these specific set of parameters ($n_e$ =3$n_c$, d = 0.75$\lambda$, L = 0.4$\lambda$), the transmitted energy fraction is substantially higher in negatively chirp pulse, the charge separation field cannot sustain for longer duration to effectively accelerate the ions. On the other hand, the number of charged particles being accelerated are almost same for all three cases. 

In Figure \ref{energy}(b), we present the accelerated ion energy spectra from the secondary layer, when the density in the PL is $3n_c$. Prominent mono-energetic ion bunches with peak energy $\mathrm{\sim22.6\,MeV}$ and energy spread ($\Delta \mathcal{E}/\mathcal{E}) \sim 28.6\%$ for the case of negatively chirped pulse was observed. 
For positively chirped the peak energy and energy spread are $\mathrm{\sim 41.7\,MeV}$ and ($\Delta \mathcal{E}/\mathcal{E}) \sim 18.29\%$. And for unchirped pulse, the peak energy and energy spread are $\mathrm{\sim 43.0\,MeV}$ and ($\Delta \mathcal{E}/\mathcal{E}) \sim 19.8\%$) respectively. 
The reason behind such mono-energetic behavior in ion spectra as reconstructed from the PIC simulation results lies in the dynamics of charge separation field.  
In Figure \ref{energy}(c), the ion energy spectra from the relativistically overdense primary layer (Deuterium) are shown, whereas in Figure \ref{energy}(d) the spectra are from  secondary layer (Hydrogen). The spectra from PL exhibits similar characteristics with Figure \ref{phasespace}(f), single layer target with the same target parameter. In Figure \ref{energy}(c), it was observed that all three cases results in the peak ion energy at the same position of almost 10 MeV, with different cut-off energy. 
The maximum ion cut-off energy for positively and negatively chirped pulses is around 30$\,$MeV, whereas, for the case unchirped pulse is around 20$\,$MeV. Since, for this density, the electrostatic field created is much lower than the case of $n_e = 3n_c$. It is evident from Figure \ref{fieldens}(d) in sec. \ref{sec:foil} for the case of unchirped pulse with single target. This lower electrostatic field, however, remains in the vicinity of the target surface. Since, laser pulse does not transmit through the target at this interaction condition, therefore, low energy ion bunches were observed from the secondary layer as shown in Figure \ref{energy}(d). The maximum ion energy of the bunches reaches upto 4$\,$MeV only for positively chirped pulse.

Therefore, we would like to highlight that either one can generate quasi-monoenergetic bunches from the single layer target in opaque region, or from secondary layer in transparency region. Additionally, we get a clear indication from the spectra about the target species being accelerated.

\section{Ion beam quality and influence of laser focusing on the acceleration regimes}
\label{sec:2d}
\begin{figure*}[ht!]
	\vskip -0.15cm
    \centering
	\includegraphics[scale=0.9]{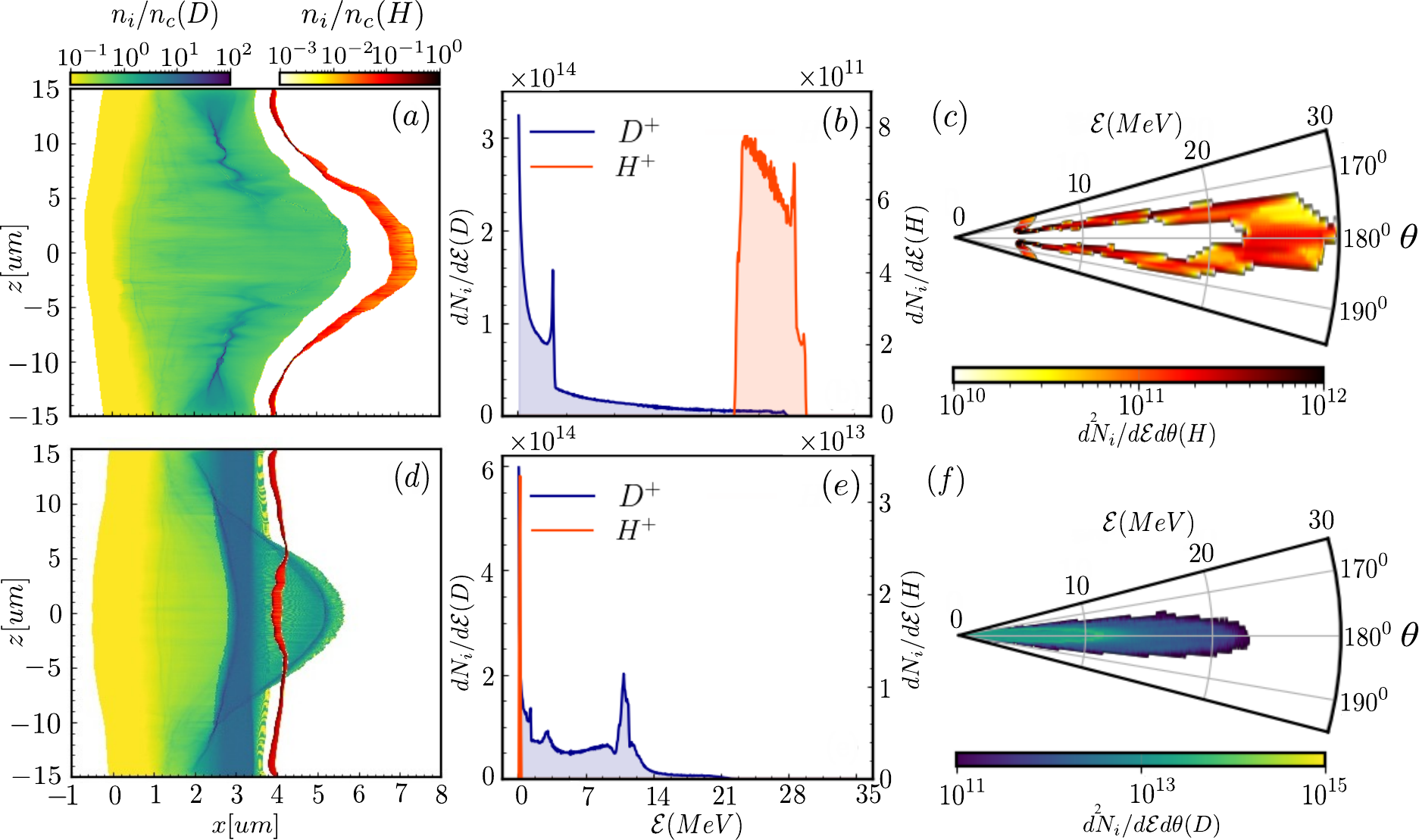}
	\vskip -0.35cm 
	\caption{ Target ion density map for the primary (Deuterium) and secondary (Hydrogen) layers are shown in (a) for $n_e = 3n_c$ (transparency regime) and (d) $n_e = 9n_c$ (reflection regime). The colorbars represent the ion densities in log-scale for both the species in yellow-green for $D^+$ and yellow-red for $H^+$. (b) and (e), ion energy spectrum corresponding to the central line outs for $n_e = 3n_c$ and $9n_c$ respectively (PL: Deuterium layer in blue color and SL: Hydrogen layer in orange color). The angular energy distribution of the predominant species is shown with the polar plots for both the cases, in (c) $\mathrm{H^{+}}$, 3$n_c$ and (f) $\mathrm{D^{+}}$, 9$n_c$. The colorbar represents the number of ions accelerated at a certain angle with a specific energy in MeV. The ion energy 2D spatial distributions and the spectra are obtained at $t = 48.806\tau$, i.e. 24$\tau$ after the peak interaction of laser pulse with the target, similar to the condition in 1D simulations.\label{2d}}
	  \vskip -0.55cm
\end{figure*}

In order to probe the interaction deeper and to validate our previous observations we have conducted 2D fully relativistic  Particle-in-Cell simulations using the code WarpX \cite{warpx}. A Gaussian laser pulse (Gaussian in both space and time) of wavelength 1$\mu m$, duration of 10.0$fs$ (FWHM of the intensity profile) as in 1D PIC simulations and beam waist ($1/e^2$ radius of intensity spatial profile at focus) of 8$\mu m$ is normally incident on a double layer plasma target. The laser and target conditions are similar to the conditions used in Figure \ref{energy}, where the primary layer (PL) target is made of Deuterium with thickness 0.75$\mu m$ and the secondary layer (SL) is made of Hydrogen having thickness of 0.2$\mu m$. The simulation box is 30$\mu m$ $\times$ 30$\mu m$, with the double layer target located at the center (x=0). The simulation runs with a spatial resolution of 170 cells per wavelength and with 4 ions and 6 electrons in each dimension, respectively. We have considered an exponential density ramp with scale-length 0.4$\mu m$ in front of the composite target. In case of PL, we have considered two electron densities $n_e = 3n_c$ and $9n_c$, with SL having $n_e =0.1n_c$.

In Figure \ref{2d}, we have presented the 2D PIC simulation results for the transparency (upper panel) and the opaque regimes (lower panel) featuring the ion density and energy spectrum from primary layer  and secondary layer for focal spot of 8$\mu m$. Figure \ref{2d} captures the time snapshot of the process, long after the interaction is over when the laser pulse has already left the simulation box either from the right side (post-transmission) or from the left side (post-reflection).
Figure \ref{2d}(a) clearly shows that the target ions are expanding with higher rate in laser propagation direction. The SL ions have moved ahead of the PL ions. As is evident, due to lower mass of ions in SL, they move faster in comparison with the ions in PL, given the same accelerating field.
Whereas, in Figure \ref{2d}(d), we observed that as the laser pulse impinges on the target, the target ions expand from both the front and back surface. This leads to the formation of double peak structure in ion density profile as in Figures \ref{ey2ez2ne}(d), \ref{fieldens}(d) and \ref{phasespace}(d). Although, ions in SL are lighter in mass, but they stay in close vicinity to primary layer ions. Since, they experience reduced charge separation field at the rear end of the target. This nature of primary layer ion density upholds with the 1D behavior, as it was observed in Figure \ref{ey2ez2ne}(b) and (d).

Next, the ion energy spectra from both the layers (PL and SL) for the relativistically transparent region ($n_e = 3n_c$) are presented in Figure \ref{2d}(b). In this case, the ions in PL exhibit the spectral signature similar to Maxwellain distribution, as in Figure \ref{energy}(a) for the case of unchirped pulse, with cut-off energy of around 20$\,$MeV. Whereas, the ion spectra from the SL are mono-energetic, peaking at $\sim$24$\,$MeV and with energy spread ($\Delta \mathcal{E}/\mathcal{E}$) of 25.9$\%$. The ion energy (peak and cut-off) obtained from 2D PIC are slightly lower than 1D PIC, however, 2D-PIC simulations are able to qualitatively reproduce the characteristic of the ion spectra. 
On the other hand, in Figure \ref{2d}(e) the ion energy spectra from overdense region ($n_e = 9n_c$) is shown. Here, from PL we observe a quasi-monoenergetic ion spectra similar to Figure \ref{energy}(b) for unchirped pulse, with peak energy at around 11$\,$MeV with energy spread ($\Delta \mathcal{E}/\mathcal{E}$) of 8.5$\%$ and cut-off energy of 22$\,$MeV. Whereas, only few  MeV ions from the secondary layer. The ion energy spectra from the overdense region ($n_e$ = 9$n_c$) closely corroborate with the 1D PIC simulation results as shown in Figure \ref{energy}(b,d).


\begin{table*}[ht!]
\small
\newcolumntype{C}{>{\centering\arraybackslash}c}
\centering 
\begin{adjustbox}{max width=1.98\columnwidth}{\begin{tabular}{|c|c|c|c|c|c|c|c|c|} \hline
\rowcolor{black} {\textcolor{white}{Peak density}} & \multicolumn{4}{c|}{\textcolor{white}{$n_e = 3n_c$}} &  \multicolumn{4}{c|}{\textcolor{white}{$n_e = 9n_c$}}\\ \hline
\rowcolor{white} Energy (MeV) & \multicolumn{2}{c|}{Peak} & \multicolumn{2}{c|}{Cut-off} &  \multicolumn{2}{c|}{Peak} & \multicolumn{2}{c|}{Cut-off} \\ \hline
\cellcolor{LightGray} Focalspot radius ($\mu$m) &PL($\mathrm{D^{+}}$)&SL($\mathrm{H^{+}}$)&PL($\mathrm{D^{+}}$)&SL($\mathrm{H^{+}}$)&PL($\mathrm{D^{+}}$)&SL($\mathrm{H^{+}}$)&PL($\mathrm{D^{+}}$)&SL($\mathrm{H^{+}}$) \\ \hline
\cellcolor{LightGray} 4& 15 & 13 & 15 & 20 & 9 & 0.73 & 21 & 1.45 \\ 
\cellcolor{LightGray} 8& 28 & 24 & 28 & 29 & 11 & 0.21 & 22 & 0.41 \\ 
\cellcolor{LightGray} 12& 41 & 38 & 42 & 45 & 10.7 & 0.21& 29 & 0.42 \\ \hline
\end{tabular}}
\end{adjustbox}
\caption{Peak and cut-off ion energy values obtained in relativistic transparent region ($n_e$ = 3$n_c$) and overdense region ($n_e$ = 9$n_c$) from double layer target configuration (primary layer (PL) as $\mathrm{D^{+}}$ and secondary layer (SL) as $\mathrm{H^{+}}$) with three different driving focal spot radius ($1/e^2$ radius of the intensity spatial envelop at focus) i.e. 4$\mu m$, 8$\mu m$ and 12$\mu m$. Figure~\ref{2d} summarizes the results corresponding to the case of intermediate focal spot radius of 8$\mu m$ presented in the table.\label{t1}}
\end{table*}
In Figure \ref{2d}(c) and (f), polar ion energy spectral map is presented from the target layers to investigate the divergence of the ion beams. Figure \ref{2d} (c) corresponds to the $\mathrm{H^+}$ ion spectrum shown in \ref{2d}(b). Whereas Figure \ref{2d} (f), corresponds to the $\mathrm{D^+}$ ion spectrum shown in \ref{2d}(e). These two cases in the simulations exhibit mono-energetic and quasi-monoenergetic spectral characteristics. Although, the divergence of ions in Figure \ref{2d}(c) is moderately higher while compared with Figure \ref{2d}(f), an interesting feature is observed in \ref{2d}(c), where ions in the central area (center of the laser focus) accelerate with higher energy compared to the ions at the peripheral area. Since, ions at the focus center experiences higher laser intensity and are pushed by the driver laser at higher rate in the transparency conditions. Therefore, only the central area of the accelerated ions in \ref{2d}(c) can construct a mono-energetic spectrum, such as shown in \ref{2d}(b). Whereas, the ion energies from the $\mathrm{D^+}$ target in opaque region show the low divergence accelerated ions, due to reflection of the driving laser pulse from the target.

For the sake of completeness and to verify the dimensional effect in our simulation, we have studied the 2D effect on ion acceleration. In 2D simulations, not only the dimension of the target changes to 2D-plane but also the driving laser pulse waist is taken into account, therefore focal spot plays a crucial role in determining the maximum or cut-off and peak ion energies from the targets. We have summarized the ion energy cut-off and peak values in Table \ref{t1} for the above mentioned target species, density and thickness with different laser focal spot sizes. We observed that in case of tight focusing i.e. 4$\mu m$, the ion energies (peak and cut-off) obtained from PL and SL in transparency region ($n_e$ = 3$n_c$) is almost half than that from focal spot size of 8$\mu m$. Similarly, if we further increase the focal spot size three-fold i.e. 12$\mu m$, the ion energies are also enhanced almost three times, when compared with focal spot size of 4$\mu m$ and almost 1.5times higher than focal spot size of 8$\mu m$. Here, we would like to specify that the laser intensity is kept constant through, thus effectively means an increase in the laser energy. On the other hand, the ion energies from the opaque region ($n_e$ = 9$n_c$) does not show significant changes with the variation in focal spot size. This is due to the completely reflection of the laser pulse irrespective to the focal spot sizes, therefore the electron dynamics and charge separation field remain almost identical for different focal spot sizes. 
When we compare these ion energies obtained from the 1D PIC simulation as in Figure \ref{energy}, are approximately in range for 12$\mu m$ focal spot size. Since, larger focal spot size (spot size $\geq$ 12$\mu m$) mimics the laser conditions as of 1D simulation near the axis.

\section{Concluding Remarks}
\label{sec:conclusions}
In this article, we have demonstrated that by tuning accessible sub-PW class laser parameters and target conditions, one can harness the transition between two-different acceleration mechanisms, namely Relativistic Induced Transparency (RIT) and Radiation Pressure Acceleration (RPA) to directly administer the ion bunch spectral features to a large extent. This study enables us to optimize the features of the accelerated ion beams by scanning the target parameters (density, thickness and geometry), but also adding additional flexibility via controlling the laser frequency chirp and prepulses (to create a suitable preplasma). We have demonstrated that the interaction conditions for each of the ion acceleration mechanisms of our interest (RIT and RPA) can be identified by analyzing the transmission energy coefficient ($\mathcal{E}_t$) with varying target and laser parameters. 

Due to very different charge particle dynamics in RIT and in RPA region, the charge separation field generated in RIT is substantially stronger when compared with the field generated in the opaque region leading to considerably different spectral features generated in RIT and RPA,  Maxwellian-like ion spectra in the RIT and quasi-mono-energetic in the RPA. We have provided a thorough prescription on how to go from one regime of operation to another and effectively generate quasimonoenergetic ion spectrun. Furthermore, we have established the proof of principal that any chosen ion species accelerated to a quasimonoenergetic energy spectrum under conditions realizable and easily controllable in experiments, using a double layer configuration. 2D PIC simulations clearly unveil the ion acceleration process from  different target layer and establish the ion beam quality. In addition, the focal spot dimension plays an important role and the ion energy spectral behavior with larger focal spot corroborate with 1D PIC simulation result. With the large number of physical parameters involved, this numerical study is important for the future experimental verification and further extension towards multidisciplinary applications of laser driven ion beams \cite{Mondal2018,DanieleM2018}. The possibility of controlling transition between different ion acceleration mechanisms, within the same experimental setup, can pave the way to optimize the ion energies and spectral features  depending on region of operation. In addition the possibility of controlled experiments in this direction might open the doors to study the correlation of ion acceleration with proposed cross disciplinary studies like strong field plasma quantum optics\cite{Lamprou2021}.

\section*{Acknowledgements}
This research has been supported by the IMPULSE project which receives funding from the European Union Framework Programme for Research and Innovation Horizon 2020 under grant agreement No 871161. ELI-ALPS is supported by the European Union and co-financed by the European Regional Development Fund (ERDF) (GINOP-2.3.6-15-2015-00001). S.K. and S.M. acknowledges Project No. 2020-1.2.4-TÉT-IPARI-2021-00018, which has been implemented with support provided by the National Research, Development and Innovation Office of Hungary, and financed under the 2020-1.2.4-TET-IPARI-CN funding scheme. S.C. and S.K. acknowledge the High Performance Computation (HPC) facility/service at ELI-ALPS.
\bibliography{ion_arXiv}

\end{document}